\documentclass[a4paper,12pt]{article}

\usepackage{graphicx}
\usepackage{amsfonts}
\usepackage{amsmath, amssymb}
\usepackage{amsthm}

\newtheorem{theorem}{Theorem}
\newtheorem{proposition}{Proposition}
\newtheorem{definition}{Definition}
\newtheorem{corollary}{Corollary}

\begin{document}

\title{Population models with singular equilibrium}

\author{Faina S Berezovskaya$^{1}$ , Artem S Novozhilov$^{2}$, Georgy P Karev$^{2,}$\footnote{Corresponding author: tel.: (301) 451-6722; e-mail: karev@ncbi.nlm.nih.gov} \\
\textit{\normalsize $^{1}$Howard University, 6-th Str., Washington DC 20059;}\\
\textit{\normalsize $^{2}$National Institutes of Health, 8600
Rockville Pike,Bethesda, MD 20894;} }

\date{}

\maketitle

\begin{abstract}

A class of models of biological population and communities with a
singular equilibrium at the origin is analyzed; it is shown that
these models can possess a dynamical regime of deterministic
extinction, which is crucially important from the biological
standpoint. This regime corresponds to the presence of a family of
homoclinics to the origin, so-called elliptic sector. The complete
analysis of possible topological structures in a neighborhood of
the origin, as well as asymptotics to orbits tending to this
point, is given. An algorithmic approach to analyze system
behavior with parameter changes is presented. The developed
methods and algorithm are applied to existing mathematical models
of biological systems. In particular, we analyze a model of
anticancer treatment with oncolytic viruses, a parasite-host
interaction model, and a model of Chagas' disease.

\end{abstract}

\section{Introduction}\label{intro}
Mathematical models of predator-prey or parasite-host interaction
and epidemiological models formulated as a system of ordinary
differential equations (ODEs) share the same framework of
modeling. The principles of these models, balance relations and
decomposition of the rates of change into birth and death
processes, which were applied since publications of Lotka-Volterra
equations \cite{Lotka,Volterra} and ODE $SIR$-model of Kermack and
McKendrick \cite{Kermack}, have remained valid until today.
Modifications were limited to replacing growth, death, and
transmission rates by more complex functions, other types of
functional responses, density-dependent mortality rates, or modes
of pathogen transmission other than mass action kinetics.

In a number of cases different population models possess a
distinct similar feature that complicates the analysis of the
models; namely, a number of models are not defined at a singular
point. Without loss of generality we can assume that this point is
the origin $O(0,0)$. The models we consider in the present paper
are formulated in such a way that this singularity is removable,
for example, we can get rid of it by a time change. After the time
change a topologically equivalent dynamical system in $\mathbb
R_+^2$ is obtained, where $\mathbb R_2^+=\{(x,y)\colon
x\geqslant0,\,y\geqslant0\}$, which is a natural state space for
biologically motivated models. The origin $O(0,0)$ becomes a well
defined equilibrium for this dynamical system. Sometimes, however,
the usual approach by linearization fails to infer the structure
of a neighborhood of $O(0,0)$ because the origin is a
non-hyperbolic equilibrium for any parameter values (i.e., this
point has both eigenvalues equal zero).

The main goal of the present paper is to describe completely the
possible structures of a small neighborhood $\Omega$ of
non-hyperbolic equilibrium point $O(0,0)$ in case of a particular
class of differential equations that includes many biological
models. We also present an algorithm to analyze such equations in
$\Omega$ and apply this algorithm to a number of mathematical
models. In particular, we show that the models under consideration
can possess a dynamical regime of deterministic extinction, which
is crucially important from the biological standpoint. This regime
corresponds to the presence of a family of homoclinics to the
origin, so-called elliptic sector. We note that the fine structure
of the phase plane containing the elliptic sector was sometimes
overlooked in the analysis of these models.

More specifically, we analyze non-hyperbolic equilibrium $O(0,0)$
of the following system of ODEs:
\begin{equation}\label{mains}
    \begin{split}
    \frac{dx}{dt} & =P_2(x,y)+P^*(x,y),\\
    \frac{dy}{dt}   & =Q_2(x,y)+Q^*(x,y),
\end{split}
\end{equation}
where $P_2(x,y),\,Q_2(x,y)$ are homogeneous polynomials of the
second order:
\begin{equation*}
    \begin{split}
    P_2(x,y)&=p_{2,1}x^2+p_{1,2}xy+p_{0,3}y^2,\\
    Q_2(x,y)&=q_{3,0}x^2+q_{2,1}xy+q_{1,2}y^2,
\end{split}
\end{equation*}
and $P^*(x,y)=O(|(x,y)|^{3}),\,Q^*(x,y)=O(|(x,y)|^{3})$, i.e.,
their Taylor series start with terms of order three or higher.

It is important to stress that many of the global properties of
the models with the discussed peculiarity are determined by the
properties of equilibrium $O(0,0)$, and, therefore, it is
essential to know the exact structure of a neighborhood of this
point and dependence of this structure on the model parameters.

We organize the paper as follows: Section 2 is devoted to examples
of biologically motivated models, which, after a suitable time
change, fall into the class of differential equations
\eqref{mains}; in Section 3 we introduce the basic definitions and
notations and state the main theorems that describe possible
topological structures of the origin of system \eqref{mains}; here
we also present an algorithm to study the structure of $\Omega$ of
$O(0,0)$; in Section 4, using the algorithm from Section 3, we
analyze some mathematical models, mainly those introduced in
Section 2; Section 5 contains discussion and conclusions; finally,
proofs of some mathematical statements are given in Appendix.

\section{Motivation and background}\label{sec:2}
\paragraph{Ratio-dependent models.}
Deterministic predator-prey models can be written in the following
`canonical' general form:
\begin{equation}\label{PP}
    \begin{split}
    \frac{dN}{dt} & =F(N)-g(N,P)P,\\
    \frac{dP}{dt}   & =eg(N,P)P-qP,
\end{split}
\end{equation}
where $N$ and $P$ are the densities (or biomasses) of prey and of
predators, respectively. The production of prey in the absence of
predators is described by the function $F(N)$, whereas $g(N,P)$ is
the functional response (number of prey eaten per predator per
unit time \cite{Holling1}). The constant $e$ is the trophic
efficiency, and predators are assumed to die with a constant death
rate $q$.

Many questions in predator-prey theory revolve around the
expression that is used for the functional response $g(N,P)$. Much
early work was only concerned with the way in which this function
varies with prey density (e.g., the so-called Holling types $I,
II, III$), ignoring the effect of predator density. For example,
in the Lotka-Volterra models, where $g(N,P)=\alpha N$, or in
models with the Holling type $II$ response \cite{Holling2}, where
$g(N,P)=\alpha N/(1+\alpha hN)$, the functional response is of the
form $g=g(N)$ (termed `prey-dependent' by Arditi and Ginzburg
\cite{Arditi3}), see also \cite{Bazykin}.

It was also recognized that the predator density could have a
direct effect on the functional response. A number of such
`predator-dependent' models have been proposed, the most widely
known are those of Hassel and Varley \cite{Hassell}, DeAngelis et
al. \cite{DeAngelis}, or Beddington \cite{Beddington}.

Arditi and Ginzburg suggested that the essential properties of
predator dependence could be rendered by a simple form which was
called `ratio-dependence'. The functional response is assumed to
depend on the single variable $N/P$ rather than on the two
separate variables $N$ and $P$. Under this supposition system
\eqref{PP} becomes
\begin{equation}\label{PPr}
    \begin{split}
    \frac{dN}{dt} & =F(N)-g(N/P)P,\\
    \frac{dP}{dt}   & =eg(N/P)P-qP.
\end{split}
\end{equation}

Discussion of the biological implications and relevance of the
ratio-dependent models, together with their principal predictions,
can be found in, e.g., \cite{Abrams,Akcakaya}.

One particular model that received considerable attention is of
the form
\begin{equation}\label{PPex}
    \begin{split}
    \frac{dN}{dt} & =rN\left(1-\frac{N}{K}\right)-\frac{\alpha NP}{P+\alpha h N}\,,\\
    \frac{dP}{dt}   & =e\frac{\alpha NP}{P+\alpha h N}-qP,
\end{split}
\end{equation}
where Michaelis-Menten or Holling type $II$ function
$$g(z)=\alpha z/(1+\alpha h z),\quad z=N/P$$
was used as the functional response.

The model \eqref{PPex} was studied in, e.g.,
\cite{Berez5,Hsu2,Jost,Kuang,Tang,Xiao}. It was shown, by this
particular example, that ratio-dependent models can display
original dynamical properties that cannot be observed in
two-dimensional prey-dependent models. For instance, the origin
can be an equilibrium point simultaneously attractive and
repelling, thus shedding light on ecological extinction.
Coexistence of several dynamical regimes with the same set of
parameters was also observed.

From the mathematical point of view, ratio-dependent predator-prey
models raise delicate questions because the functional response is
undefined at the origin $N=0,\, P=0$. As a consequence, the origin
is a so-called non-analytical complicated equilibrium point
\cite{Berez5}.

We are only interested in the dynamics of system \eqref{PPr} in
$\mathbb R_2^+=\{(N,P)\colon N\geqslant0,\,P\geqslant0\}$. Time
scale change $dt\rightarrow (P+\alpha h N)dt$ for system
\eqref{PPex} results in the system
\begin{equation}\label{PPexP}
    \begin{split}
    \frac{dN}{dt} & =rN(1-N/K)(P+\alpha h N)-\alpha NP,\\
    \frac{dP}{dt}   & =e\alpha NP-qP(P+\alpha h N).
\end{split}
\end{equation}
The system \eqref{PPexP} is well defined at $O(0,0)$, however, the
Jacobian matrix at $O(0,0)$ is a zero matrix, i.e., $O(0,0)$ is a
non-hyperbolic equilibrium of \eqref{PPexP}. Obviously, model
\eqref{PPexP} belongs to the class of ODEs \eqref{mains}.

\paragraph{Frequency-dependent pathogen transmission in a population of variable size.}
Transmission is the key process in a host-pathogen interaction. In
most early models transmission was assumed to occur through the
law of mass-action. For example, in the classical model of two
differential equations, formulated as a particular case of the
general model of Kermack and McKendrick in their pioneer work
\cite{Kermack}, the transmission function takes the form $\beta
SI$. Here $S$ is the density of susceptible individuals, $I$ is
the density of infective individuals, and $\beta$ is the
transmission coefficient. Note that, for mass action, the contacts
per unit time per individual rise linear with the population
density $N=S+I$ (careful elaboration on the models of contact
process can be found in \cite{Diekmann}).

At the other extreme, the contact rate might be independent of
host density. Assuming that susceptible and infective were
randomly mixed, this would lead to transmission following $\beta
SI/N=\beta SI/(S+I)$. This mode of transmission is often called
`frequency-dependent' or `density-independent' transmission
\cite{McCallum}. It is often assumed in models of sexually
transmitted diseases because the number of sexual partners of an
individual usually depends on the mating system of the species and
is weakly related to host density \cite{May}.

We note here that the distinction between these two modes of
transmission is not crucially important for many problems in human
diseases, because most pathogens cause little mortality and the
total population size remains more or less constant
\cite{Hethcote}.

Even if some demography processes are taken into account many
epidemiological models are formulated so that the infectious
disease spreads in a population that has a fixed size
$N=\textrm{const}$ with balancing inflows and outflows due to
birth or death or migration, see, e.g.,
\cite{Bussenberg1,Hethcote}. However, if the population growth or
decrease is significant (it might be the case on long periods of
time, e.g., for diseases transmitted vertically) or the disease
causes enough death to influence the population size, then it is
not reasonable to assume that the population size is constant. In
this case the difference between the mass action transmission and
the frequency-dependent transmission becomes profound
\cite{Gao,Menalorca}. It is worth noting that the
frequency-dependent transmission is considered to be more
realistic for most human diseases \cite{Hethcote}.

Due to the particular form of the transmission function, $\beta
SI/(S+I)$, the origin $S=0,\,I=0$ becomes undefined in models
where host extinction is a possibility, and the properties of the
origin crucially affect the global dynamics of the system.

As an example of such models, we rewrite a model of Chagas'
disease from \cite{Bussenberg1,Bussenberg2} that takes the form
\begin{equation}\label{SIex1}
\begin{split}
   \frac{dS}{dt} &=(b-r-v)S+(b_1(1-q)+c)I-\beta\frac{SI}{S+I}\,,\\
   \frac{dI}{dt} &=(b_1q-r_1-c)I+vS+\beta\frac{SI}{S+I}\,,
\end{split}
\end{equation}
where $b,\,b_1,\,r,\,r_1,\,c,\,q$, respectively, denote the birth
rates of susceptible and infective individuals, their death rates,
the cure rate and the probability of vertical transmission; $v$ is
the transmission rate of disease by a vector. From \eqref{SIex1}
we have
$$
\frac{d}{dt}(S+I)=(b-r)S+(b_1-r_1)I,
$$
and the total population could be either increasing or decreasing.

A usual way to analyze models like \eqref{SIex1} is to perform the
change of variables $x=S/(S+I),\,y=I/(S+I)$. After this and using
the fact that $x+y=1$, it is often straightforward to obtain the
analysis of the system in coordinates $x,\,y$, simultaneously
keeping track of the asymptotic behavior of the total population
size $N$ \cite{Bussenberg1,Gao,Menalorca}. However, after this
transformation some of the information concerning initial
variables $S$ and $I$ can be lost. For example, in variables $x$
and $y$ it is difficult to see that the origin can be
simultaneously attractive and repelling, and deterministic
extinction of the total population can be accompanied by an
initial growth of the susceptible subpopulation (see Section
\ref{sec:4}, Example 3).

Though mathematical models of pathogen transmission with function
$\beta SI/N$ were extensively studied for years, the first case
study, to our knowledge, that focuses on the possibility of
deterministic extinction of host population was considered in
\cite{Hwang1}. A careful analysis of a more general model was
presented in \cite{Berez6}.

As an example, let us state the model of host-parasite interaction
from \cite{Hwang2}. This model allows for host extinction and,
formulated through the basic birth and death processes, has the
form
\begin{equation}\label{SIex}
    \begin{split}
    \frac{dS}{dt} & =a(S+I)-a(1-\theta)I-cS(S+I)-(d+m)S-\beta\frac{SI}{S+I}\,,\\
    \frac{dI}{dt}   & =-(d+\alpha)I-cI(S+I)+\beta\frac{SI}{S+I}\,,
\end{split}
\end{equation}
where $a,\,c,\,d,\,m,\,\alpha$ are nonnegative parameters, and
$0\leqslant\theta\leqslant1$. Model \eqref{SIex} was reduced to a
Gause-type system by means of one blow-up transformation
$(S,I)\rightarrow (u,I),\,u=S/I$. We show (Section \ref{sec:4},
Example 2) that using this transformation is not enough to obtain
a complete qualitative picture of a neighborhood of $O(0,0)$.

Systems \eqref{SIex1} and \eqref{SIex} can be transformed into
polynomial systems with a non-hyperbolic equilibrium at the origin
by the time change $dt\rightarrow (S+I)dt$, and the resulting
systems of ODEs are in the class \eqref{mains}.

It is worth mentioning that if the pathogen transmission follows
mass action kinetics, but $S$ and $I$ denote numbers and not
densities, and the total population density (here we speak of
local spatial density) remains constant, one has to use $\beta
SI/(S+I)$ expression to model transmission of disease
\cite{deJong}.

\paragraph{Other models.}
We by no means gave a full list of models analysis of which
requires the knowledge of the structure of $\Omega$ of undefined
equilibrium $O(0,0)$ and which can be reduced to \eqref{mains}.
Similar models appear in many other areas of mathematical modeling
in biology. Some mathematical models of interaction between
populations of cells with a virus population can also be
formulated in the form \eqref{mains}; e.g., one such model was
applied to simulate anticancer therapy with oncolytic viruses
\cite{Novozhilov} (see Section \ref{sec:4}, Example 1). It was
shown that the model possesses dynamical regimes that lead to
elimination of cancer cells (a phenomenon known from clinical
trials). Recently it was suggested that immune system response is
more consistent with empirical data if it is considered to be
ratio-dependent, see, e.g., \cite{dePillis2}.


\section{Non-hyperbolic equilibrium of system \eqref{mains} and structures of its neighborhood}\label{sec:3}
\subsection{Preliminaries}

Let $O(0,0)$ be an isolated singular point of the vector field
\begin{equation}\label{3:0}
    J(x,y)=P(x,y)\frac{\partial}{\partial x}+Q(x,y)\frac{\partial}{\partial y}
\end{equation}
and, correspondingly, an isolated equilibrium of the system of
ODEs
\begin{equation}\label{3:1}
    \frac{dx}{dt}=P(x,y),\quad\frac{dy}{dt}=Q(x,y).
\end{equation}
Here
\begin{equation*}
    P(x,y) =P_n(x,y)+P^*(x,y),\quad     Q(x,y)   =Q_n(x,y)+Q^*(x,y),
\end{equation*}
where $P_n(x,y),\,Q_n(x,y)$ are homogeneous polynomials of the
$n$-th order:
\begin{equation*}
    \begin{split}
    P_n(x,y)&=p_{n,1}x^n+p_{n-1,2}x^{n-1}y+\ldots+p_{0,n+1}y^n,\\
    Q_n(x,y)&=q_{n+1,0}x^n+q_{n,1}x^{n-1}y+\ldots+q_{1,n}y^n,
\end{split}
\end{equation*}
and $P^*(x,y)=O(|(x,y)|^{n+1}),\,Q^*(x,y)=O(|(x,y)|^{n+1})$.

It is well known (e.g., \cite{Andronov}) that point $O(0,0)$ can
be either \textit{monodromic} (i.e., a focus or a center) or all
orbits of system \eqref{3:1}, tending to the origin for
$t\to\infty$ or $t\to-\infty$ (we shall call such orbits as
\textit{$O$-orbits}), approach the origin along
\textit{characteristic directions}. The latter means that any
$O$-orbit has definite direction $\theta$ so that $\arctan
(y(t)/x(t))\to\theta$ for $t\to\infty$ or $t\to-\infty$. We note
that the number of characteristic directions is finite in the
generic case. Hereafter we refer to Andronov et al.
\cite{Andronov} for a number of results and notations used, but
restate some of the results in the form convenient for our
exposition.

Following the results from Ch.VIII Sec. 17 of \cite{Andronov} we
can state that any small enough neighborhood $\Omega$ of an
isolated equilibrium point of an analytic system can be
partitioned by $O$-orbits with characteristic directions into open
regions, called \textit{sectors}. These sectors can be classified
into three types: \textit{parabolic sectors}, \textit{hyperbolic
sectors}, and \textit{elliptic sectors}, respectively; we shall
call them \textit{the Brouwer sectors}. The Brouwer sectors are
described in the following figure (Fig. \ref{fig:1}) and their
definitions are given, e.g., in \cite{Andronov}. Note that the
topological equivalence of a sector to one of the sectors in Fig.
\ref{fig:1} need not preserve the directions of the flow.
\begin{figure}
\centering
\includegraphics[width=0.9\textwidth]{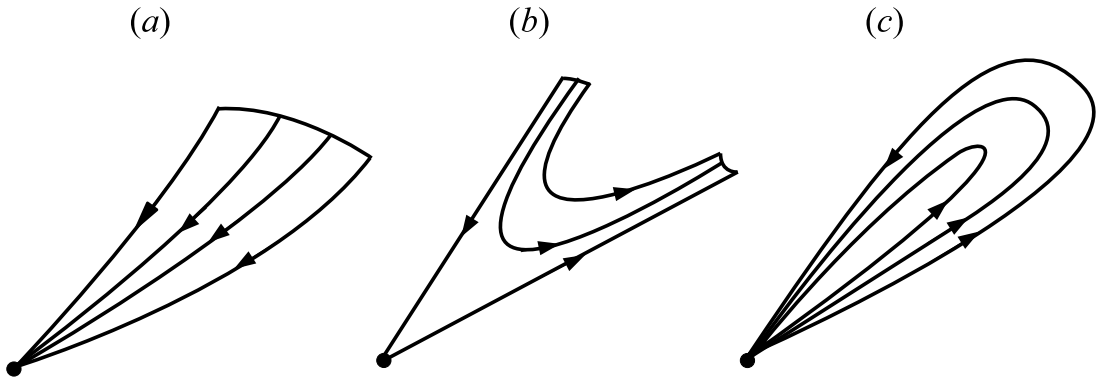}
\caption{The Brouwer sectors. (a) A parabolic sector\,; (b) a
hyperbolic sector\,; (c) an elliptic sector}
\label{fig:1}       
\end{figure}

More exactly, the next statement is valid.
\begin{theorem}\label{th3:1} There exists a neighborhood $\Omega$ of a non-monodromic singular point
of $C^\infty$-vector field, such that any neighborhood
$\tilde{\Omega}\subset\Omega$ that contains this singular point
can be partitioned into a finite number of sectors of parabolic,
hyperbolic and elliptic types. The number, types and the cyclic
order of a position of these sectors hold with decreasing of the
size of neighborhood $\tilde{\Omega}$.
\end{theorem}

For $n=1$ we have
$$
P_1(x,y)=p_{11}x+p_{02}y\,,\quad Q_1(x,y)=q_{20}x+q_{11}y\,.
$$
Equilibrium $O(0,0)$ of \eqref{3:1} with $n=1$ is hyperbolic if
$D=p_{11}q_{11}+p_{02}q_{20}\neq 0$ and $Tr=p_{11}+q_{11}\neq0$
for $D>0$. Point $O(0,0)$ has $O$-orbits with characteristic
directions if $Tr^2-4D\geqslant0$. If the last condition holds and
additionally $D\neq 0$ then $O(0,0)$ is a saddle whose
neighborhood $\Omega$ contains four hyperbolic sectors, or
$O(0,0)$ is a node whose neighborhood $\Omega$ contains only
parabolic sectors. In general, the type of the origin of system
\eqref{3:1} in the case $n=1$ is determined by the calculation of
eigenvalues, and there is a simple algorithm to infer the
structure of a small neighborhood of this point.

For $n\geqslant2$ the situation is more complicated. Equilibrium
point $O(0,0)$ is not hyperbolic (both eigenvalues are zero). The
following problem is of principal interest: \textit{to describe
all possible topological structures of a small neighborhood
$\Omega$ of $O(0,0)$ and asymptotic behavior of orbits with
variation of the system coefficients}. In the present work we
solve this problem in an important case $n=2$ under some
additional assumptions on \eqref{3:1}.

The problem of finding and describing the sequence of the Brouwer
sectors that constitute $\Omega$, as well as asymptotics of
$O$-orbits, was studied in a number of works
\cite{Berez1,Berez2,Berez3,Berez4,{Bruno},{Dumortier}}. The main
attention was paid to constructing algorithms of analysis of a
non-hyperbolic equilibrium with the parameter values fixed. The
general blow-up method, developed in
\cite{{Bruno},{Dumortier},Berez3} for so-called `system with a
fixed Newton diagram', allowed to describe topological structures
of $\Omega$ and compute asymptotics of $O$-orbits close to the
singular point. Below we apply the developed methods to system
\eqref{3:1} with $n=2$ and present an approach to analyze the
behavior of \eqref{3:1} with parameter changes.

\subsection{Basic theorems}
Let us consider system \eqref{3:1} with $n=2$ in neighborhood
$\Omega$ of an isolated equilibrium $O(0,0)$, and let
$$F(x,y)=xQ_n(x,y)-yP_n(x,y).$$
\begin{definition}
We shall call vector field \eqref{3:0} \emph{(}and the
corresponding system of ODEs \eqref{3:1}\emph{)} non-degenerate in
$\Omega$ if
\begin{equation*}
\begin{split}
   (C1)\qquad &\mbox{polynomials $P_n(x,y),\,Q_n(x,y)$ have no common factors of the form}\\
&\mbox{$ax+by$, where at least one the constants $a,\,b$ is
non-zero;}\\
    (C2)\qquad &\mbox{polynomial $F(x,y)$ has no factors of the form $(ax+by)^k$, where $k>1$.}
\end{split}
\end{equation*}
\end{definition}

\begin{figure}[thb]
\centering
\includegraphics{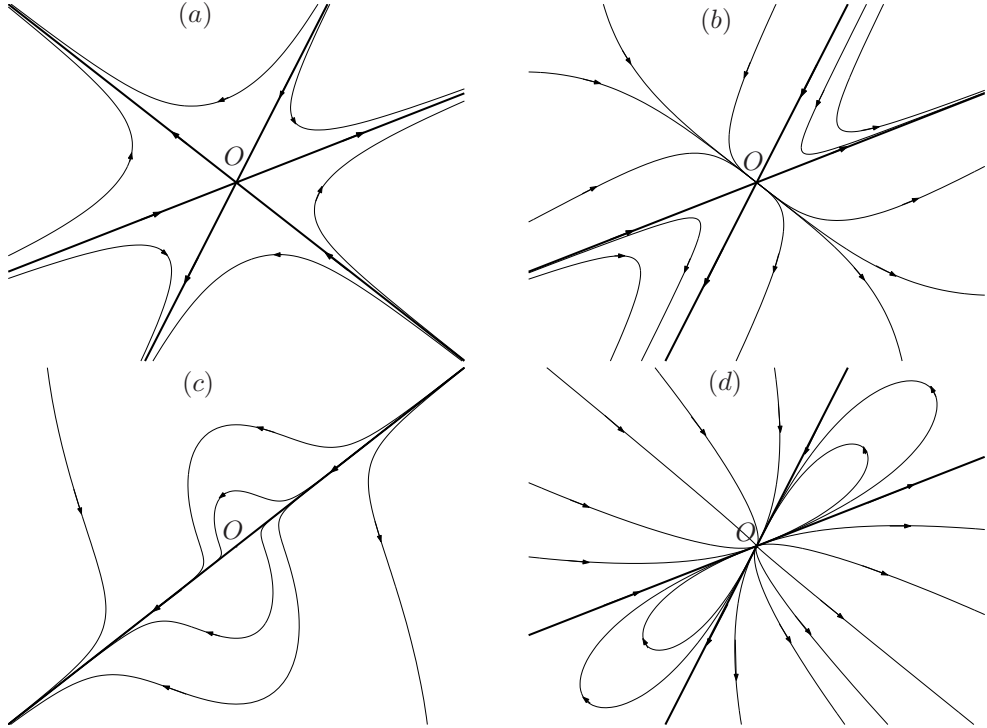}
\caption{Four possible types of the topological structure of
$\Omega$ of $O(0,0)$ for system \eqref{3:1} with $n=2$}
\label{fig:2}       
\end{figure}

The coefficients of polynomials $P_2(x,y)$ and $Q_2(x,y)$ can be
considered as the system parameters, and the parameter space is
divided into domains of topologically equivalent behavior. The
main results of our analysis show that each parameter domain
corresponds to one of the following four types of phase portraits
(Fig. \ref{fig:2}).

\begin{theorem}\label{th3:2}
Let system \eqref{3:1} with $n=2$ be non-degenerate. The
positional relationship of the Brouwer sectors in $\Omega$ can be
of four topologically non-equivalent cases:

\emph{{(i)}} six hyperbolic sectors \emph{(}Fig.
\ref{fig:2}a\emph{)};

\emph{{(ii)}} two hyperbolic sectors separated by two parabolic
sectors such that one of the parabolic sectors is attracting (in a
sense that $O$-orbits tend to $O(0,0)$ for $t\to\infty$) and
another is repelling ($O$-orbits tend to $O(0,0)$ for
$t\to-\infty$) \emph{(}Fig. \ref{fig:2}b\emph{)};

\emph{(iii)} two hyperbolic sectors \emph{\emph{(}}Fig.
\ref{fig:2}c\emph{)};

\emph{(iv)} two elliptic sectors separated by two parabolic
sectors such that one of the parabolic sectors is attracting and
another is repelling \emph{(}Fig. \ref{fig:2}d\emph{)}.
\end{theorem}

\textbf{Remarks to Theorem \ref{th3:2}.}

1. In the  case considered there are always $O$-orbits with
characteristic directions.

2. The boundaries between the parameter domains are given by
violations of conditions $(C1)$ and $(C2)$. Violation of $(C1)$
leads to the presence of a straight line, passing through the
origin, of non-isolated equilibrium points of \eqref{3:1},
violation of $(C2)$ may lead to disappearance of characteristic
directions.

3. The topological equivalence to one of the presented in Fig.
\ref{fig:2} cases need not preserve the directions of the flow.

\smallskip

To describe possible asymptotics of $O$-orbits of system
\eqref{3:1} we require the following condition:
$$
(C3)\qquad P_n(0,y)\equiv0\Rightarrow P^*(0,y)\equiv 0,\quad
Q_n(x,0)\equiv0\Rightarrow Q^*(x,0)\equiv 0.
$$

Recall that curve $y=f(x)$, $f(0)=0$ has \textit{an exponential
asymptotic} with positive power $\rho$ and non-zero coefficient
$k$ if
\begin{equation}\label{3:2}
    f(x)=kx^{\rho}(1+o(1)),\quad k\neq 0,\,\rho>0.
\end{equation}

\begin{theorem}\label{th3:3}
Let system \eqref{3:1} be non-degenerate. Then

\emph{(i)} An $O$-orbit with a characteristic direction has
asymptotic \eqref{3:2} with $\rho=1$ if and only if polynomial
$F(1,u)$ has a root $\hat{u}=k\neq0$; if $P_2(0,y)Q_2(x,0)\neq0$
then system \eqref{3:1} has no $O$-orbits with other asymptotics;

\emph{(ii)} If $Q_2(x,0)\equiv0$ and $(C3)$ holds then system
\eqref{3:1} has trivial orbits $y=0$ for $x>0$ and $x<0$; if
additionally $\beta=q_{2,1}/p_{2,1}>1$ then \eqref{3:1} has a
family of exponential asymptotics \eqref{3:2} with $\rho=\beta$,
and $k\neq 0$ is an arbitrary constant;

\emph{(iii)} If $P_2(0,y)\equiv0$ and $(C3)$ holds then system
\eqref{3:1} has trivial orbits $x=0$ for $y>0$ and $y<0$; if
additionally $\beta=p_{1,2}/q_{1,2}>1$ then \eqref{3:1} has a
family of exponential asymptotics \eqref{3:2} with $\rho=1/\beta$,
and $k\neq 0$ is an arbitrary constant.
\end{theorem}

The proofs of Theorems \ref{th3:2} and \ref{th3:3} are given in
Appendix.
\subsection{An algorithmic approach to the analysis of the structure of the origin of system
\eqref{mains}}\label{sec3:3}

The results from the previous section show \textit{what} can be
expected in neighborhood $\Omega$ of the origin of system
\eqref{mains}. Here we summarize a practical recipe to construct a
phase-parameter portrait of \eqref{mains}, i.e., we show
\textit{how} to analyze a particular system. The presented
algorithm follows directly from the proof of Theorem \ref{th3:2}
(see Appendix).

Due to the homogeneous form of system \eqref{mains} the analysis
of an isolated equilibrium $O(0,0)$ can be formulated in terms of
functions depending on one variable. Let
$F(x,y)=xQ_2(x,y)-yP_2(x,y)$. We need to consider two pairs of
polynomials
$$
\textrm{(i)}\qquad P_2(1,u),\quad F_1(u)=F(1,u),
$$
and
$$
\textrm{(ii)}\qquad Q_2(v,1),\quad F_2(v)=-F(v,1).
$$
The conditions of non-degeneracy $(C1)$ and $(C2)$ take the form:

$(C1')$ The pair of polynomials $P_2(1,u),\,Q_2(1,u)$, as well as
the pair of polynomials $P_2(v,1),\,Q_2(v,1)$, have no common
roots;

$(C2')$ The polynomials $F_1(u)$ and $F_2(v)$ have no multiple
roots.

The analysis relies on finding the roots of $F_1(u)$ and $F_2(v)$.
There can be five different cases, where we list only the roots
that are necessary for the subsequent calculations:
\smallskip

(i) $F_1(u)$ has three real roots $\hat{u}_i$, $i=1,2,3$;

(ii) $F_1(u)$ has two real roots $\hat{u}_i\neq 0$, $i=1,2,$ and
zero is a
 root of $F_2(v)$;

(iii) $F_1(u)$ has two real roots $\hat{u}_1=0,\,\hat{u}_2\neq 0$,
 and zero is a root of $F_2(v)$;

(iv) $F_1(u)$ has one real root;

(v) $F_1(u)$ has no real roots, and zero is a root of $F_2(v)$.
\smallskip

The lines $y=\hat{u}_ix$ and, if zero is a root of $F_2(v)$, $x=0$
divide $\Omega$ of $O(0,0)$ into six (cases (i)-(iii)) or two
(cases (iv)-(v)) sectors with six or two branches. These sectors
can be of a hyperbolic, elliptic, or parabolic type. Each line
consists of two branches divided by the point $O(0,0)$.

We introduce the following notations. For the branches of lines
$y=\hat{u}_ix$ we consider
$$\lambda_1^u(\hat{u}_i)=P_2(1,\hat{u}_i),\quad\lambda_2^u(\hat{u}_i)=F_1'(\hat{u}_i),$$ for the branches of line
$x=\hat{v}=0$ we consider
$$\lambda_1^v(\hat{v})=Q_2(\hat{v},1),\quad\lambda_2^v(\hat{v})=F_2'(\hat{v}).$$ Note that we need to use
functions $F_2(v)$ and $Q_2(v,1)$ only in cases (ii), (iii) and
(v).

We shall call numbers $\lambda_1^u,\,\lambda_2^u,\,\lambda_1^v$
and $\lambda_2^v$ \textit{the branch characteristics}.

\begin{definition} We shall call a branch of a sector in the phase
space $(x,y)$ of system \eqref{mains} hyperbolic if the inequality
$\lambda_1^u\lambda_2^u<0$ \emph{(}or
$\lambda_1^v\lambda_2^v<0$\emph{)} holds, and parabolic if
$\lambda_1^u\lambda_2^u>0$ \emph{(}or
$\lambda_1^v\lambda_2^v>0$\emph{)} holds, where
$\lambda_1^u,\,\lambda_2^u$ \emph{(}or
$\lambda_1^v,\,\lambda_2^v$\emph{)} the branch characteristics.
\end{definition}

Let $V$ be a sector in the state space of system \eqref{mains}
with a vertex at $O(0,0)$ composed by branches of the lines
$y=\hat{u}_1x$ and $y=\hat{u}_2x$ (or one of the lines can be
$x=0$), and $V_{\Omega}=V\bigcap\Omega$. The following
proposition, proved in Appendix, holds.

\begin{figure}
\centering
\includegraphics[width=0.9\textwidth]{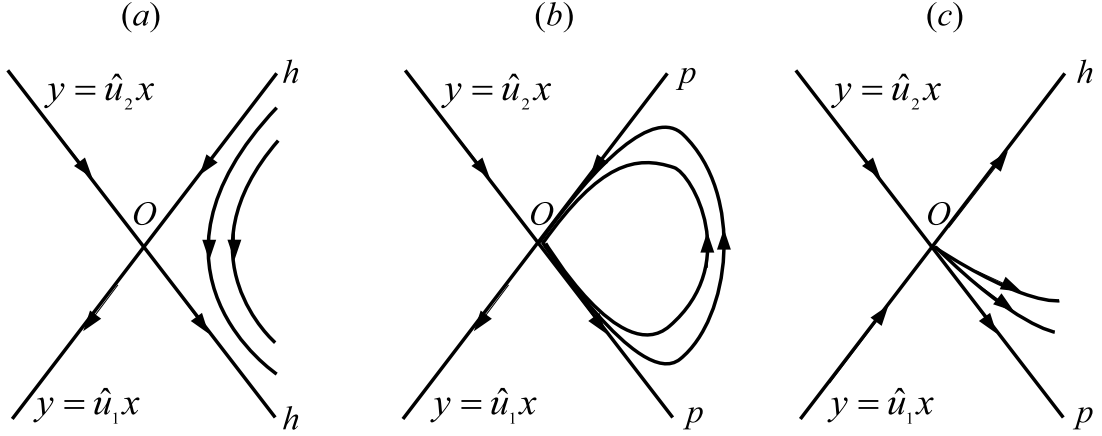}
\caption{Placing of the Brouwer sectors depending on the type of
the sector branches ($h$ -- hyperbolic branch, $p$ -- parabolic
branch)}
\label{fig:3}       
\end{figure}

\begin{proposition}\label{pr}
$V_{\Omega}$ contains

\emph{(i)} a hyperbolic sector if both of its branches are
hyperbolic \emph{(}Fig. \ref{fig:3}a\emph{)};

\emph{(ii)} an elliptic sector if both of its branches are
parabolic \emph{(}Fig. \ref{fig:3}b\emph{)};

\emph{(iii)} a parabolic sector \emph{(}or its part\emph{)} if one
of the branches is hyperbolic and another is parabolic
\emph{(}Fig. \ref{fig:3}c\emph{)}.
\end{proposition}

The direction of the phase flow on the sector branches is
determined by the sign of $\lambda_1$. If
$\lambda_1^u(\hat{u}_i)<0$ then the phase flow goes to $O(0,0)$
for $x>0$ and from $O(0,0)$ for $x<0$ (see, e.g., the line
$y=\hat{u}_1x$ in Fig. \ref{fig:3}a); if
$\lambda_1^u(\hat{u}_i)>0$ then the phase flow goes from $O(0,0)$
for $x>0$ and to $O(0,0)$ for $x<0$ (the line $y=\hat{u}_2x$ in
Fig. \ref{fig:3}a). If $\lambda_1^v(0)<0$ then the phase flow goes
to $O(0,0)$ for $y>0$ and from $O(0,0)$ for $y<0$; if
$\lambda_1^v(0)>0$ then the phase flow goes from $O(0,0)$ for
$y>0$ and to $O(0,0)$ for $y<0$.

The conditions that determine the boundaries of topologically
non-equivalent domains in the parameter space now can be
reformulated in terms of the branch characteristics. Namely,
$$\lambda_1^u(\hat{u}_i)\lambda_2^u(\hat{u}_i)=0$$ and, if 0 is a
root of $F_2(v)$, $$\lambda_1^v(0)\lambda_2^v(0)=0.$$ These
conditions are more convenient for most practical purposes.

\section{Examples}\label{sec:4}
In this section we give a number of examples of analysis of the
complicated non-analytical equilibrium point in biological models.
Mainly we restrict our attention to $\mathbb R_+^2$, but emphasize
that in some cases it is necessary to analyze not only the
behavior of the state variables in $\mathbb R_+^2$ but also the
behavior in adjacent areas.

\paragraph{Example 1.}(\textit{Mathematical model of anticancer treatment with oncolytic
viruses}) Oncolytic viruses that specifically target tumor cells
are promising new therapeutic agents \cite{Parato}. The
interaction between an oncolytic virus and tumor cells is highly
complex and nonlinear. Hence, to precisely define the conditions
that are required for successful therapy by this approach,
mathematical models are needed. Our model \cite{Novozhilov} was
formulated through the incorporation of frequency-dependent mode
of virus transmission into the model of Wodarz \cite{Wodarz1}. The
model, which considers two types of tumor cells growing in
logistic fashion, has the following form (non-dimensional
variables and parameters are used):
\begin{equation}\label{ex2:1}
    \begin{split}
    \frac{dx}{dt} & =x(1-(x+y))-\beta \frac{xy}{x+y}\,,\\
      \frac{dy}{dt}  &= \gamma y(1-(x+y))+\beta \frac{xy}{x+y}-\delta y.
\end{split}
\end{equation}
Here $x,\,y$ are (scaled) sizes of non-infected and infected tumor
cell populations, respectively; $\gamma,\,\beta,\,\delta>0$ are
non-dimensional parameters ($\beta$ takes stock of the
transmission rate, and $\delta$ describes the virus cytotoxicity).

\begin{figure}
\centering
\includegraphics[width=0.65\textwidth]{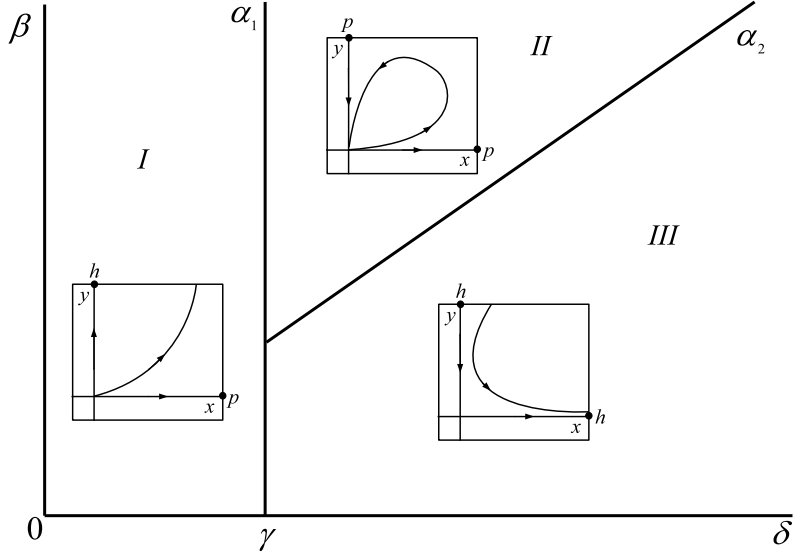}
\caption{Phase-parameter portrait of a neighborhood of the origin
of system \eqref{ex2:1} given as a cut of positive parameter space
for an arbitrary fixed value of $\gamma>0$. The bifurcation
boundaries are $\alpha_1=\{(\delta,\gamma)\colon \delta=\gamma\}$,
and $\alpha_2=\{(\delta,\gamma)\colon \beta=\delta+1-\gamma\}$
($h$ -- hyperbolic branch, $p$ -- parabolic branch)}
\label{fig:ex2}       
\end{figure}

After the time change $dt\to(x+y)dt$ we obtain the system of ODEs
in the form \eqref{mains}. We have
$$
P_2(x,y)=x^2+(1-\beta)xy,\quad
Q_2(x,y)=(\gamma-\delta+\beta)xy+(\gamma-\delta)y^2.
$$
The polynomial
$$
F_1(u)=Q_2(1,u)-uP_2(1,u)=(\gamma-\delta+\beta-1)u(u+1),
$$
if $\gamma-\delta+\beta\neq 1$, has two roots $\hat{u}_1=0$ and
$\hat{u}_2=-1$. Polynomial $F_2(v)$ also has two roots, one of
which $\hat{v}=0$ (another one is $\hat{v}_2=1/\hat{u}_2$ and we
do not need to consider it). $\mathbb R_+^2$ is confined by the
branches of the lines $y=\hat{u}_1x=0$ and $x=\hat{v}=0$. This
means that the root $\hat{u}_2$ is redundant for our analysis as
far as we are concerned with the system behavior in $\mathbb
R_+^2$. The branch characteristics are
$$
\lambda_1^u(0)=1,\quad \lambda_2^u(0)=\gamma-\delta+\beta-1,
$$
and
$$
\lambda_1^v(0)=\gamma-\delta,\quad \lambda_2^v(0)=-\lambda_2^u(0).
$$
Analyzing the sign of these expressions allows us to completely
describe a small neighborhood of $O(0,0)$ (Fig. \ref{fig:ex2} and
the following theorem).

\begin{theorem}\label{th_ex2}
For different positive values of parameters $\delta,\,\beta$, and
$\gamma$ there exist three types of topologically different
structures of the neighborhood $\Omega_+=\Omega\bigcap\mathbb
R_+^2$ of point $O(0,0)$ \emph{(}and, correspondingly, three
topologically different phase portraits of system
\eqref{ex2:1}\emph{)}:

\emph{(i)} a repelling parabolic sector \emph{(}domain $I$ in Fig.
\ref{fig:ex2}\emph{)} for the parameter values  $\delta<\gamma$.
The phase curves of the system that tend to $O(0,0)$ are of the
form
\begin{equation}\label{ex2:2}
    y=Cx^{\gamma+\delta-\beta}(1+o(1)),
\end{equation}
if $\beta>\delta+1-\gamma$, and
\begin{equation}\label{ex2:3}
    y=Cx^{(\gamma-\delta)/(1-\beta)}(1+o(1)),
\end{equation}
if $\beta<\delta+1-\gamma$; here $C\neq 0$ is an arbitrary
constant;

\emph{(ii)} an elliptic sector \emph{(}domain $II$ in Fig.
\ref{fig:ex2}\emph{)} composed by trajectories tending to $O(0,0)$
as $t\to\infty$ \emph{(}with asymptotics given by
\eqref{ex2:3}\emph{)}, as well as with $t\to-\infty$ \emph{(}with
asymptotics given by \eqref{ex2:2}\emph{)} if $\delta>\gamma$ and
$\beta>\delta+1-\gamma$;

\emph{(iii)} a saddle sector \emph{(}domain $III$ in Fig.
\ref{fig:ex2}\emph{)} for the parameter values $\delta>\gamma$ and
$\beta<\delta+1-\gamma$.
\end{theorem}

\textbf{Remark to Theorem \ref{th_ex2}.} Note that only one of two
possible arrangements of the sector branches is shown in domain
$I$ of Fig. \ref{fig:ex2}. It is possible that the branch
$x=0,\,y>0$ may be parabolic, and the branch $y=0,\,x>0$ may be
hyperbolic. In both cases, though, we obtain that $O(0,0)$ is a
repelling node, and $\Omega_+$ contains a parabolic sector.
\smallskip

The most beneficial domain of parameter values is domain $II$ in
Fig. \ref{fig:ex2} (an elliptic sector). This domain corresponds
to the total elimination of both cell populations (infected and
uninfected) regardless of the initial conditions (this dynamical
regime is observed in experimental studies, e.g.,
\cite{Harrison}). However, on its way to extinction, the overall
tumor size $x+y$ can reach rather high values (which, with the
parameters fixed, crucially depend on the initial conditions).
This indicates that we must not only identify the conditions that
favor tumor elimination, but also develop the optimal strategy to
infect the initial tumor. In the framework of the considered model
this can be done using the information on asymptotics
\eqref{ex2:2} and \eqref{ex2:3} (see \cite{Novozhilov}).

\paragraph{Example 2.}(\textit{Parasite-host interaction model}) Hwang and Kuang \cite{Hwang2}
formulated a host-parasite model allowing for host extinction. The
model, which is a non-dimensional version of \eqref{SIex}, is of
the form
\begin{equation}\label{ex1:1}
    \begin{split}
    \frac{dx}{dt} & =x+y-(1-\theta)y-x(x+y)-\delta x-s \frac{xy}{x+y}\,,\\
    \frac{dy}{dt}   & =-(\delta+r)y-y(x+y)+s\frac{xy}{x+y}\,,
\end{split}
\end{equation}
where all parameters are non-negative, and
$0\leqslant\theta\leqslant1$. Solutions of \eqref{ex1:1} are
considered in $\mathbb R^2_+$. Using the transformation $v=x/y$,
Hwang and Kuang reduced model \eqref{ex1:1} to a Gause-type system
which was completely studied.

After the time change $dt\to(x+y)dt$ we obtain
\begin{equation*}\label{ex1:2}
    \begin{split}
P_2(x,y)&=(1-\delta)x^2+(1-\delta-s+\theta)xy+\theta y^2,\\
Q_2(x,y)&=(s-\delta-r)xy-(\delta+r)y^2,\\
    F_1(u) & =\left( -r+s-1 \right) u+ \left( -r-1+s-\theta \right) {u}^{2}-\theta
\,{u}^{3}.
\end{split}
\end{equation*}

The polynomial $F_1(u)$ has three roots
$$
\hat{u}_1=0,\quad\hat{u}_2=-1,\quad\hat{u}_3=(s-r-1)/\theta,
$$
if $\hat{u}_3\neq 0,\,-1$ (case (i) in Section \ref{sec3:3}).

We note that if $\delta>1$ then, from \eqref{ex1:1},
$\frac{d}{dt}(x+y)<0$ follows. Thus, we shall consider only the
case when $0<\delta<1$.

The state space $\mathbb R_+^2$ may be contained in two sectors,
so in this case we need to calculate the branch characteristics
for the branches of three lines $y=\hat{u}_ix$. We obtain
$$
\lambda_1^u(\hat{u}_1)=1-\delta,\quad
\lambda_2^u(\hat{u}_1)=s-r-1,\quad\lambda_1^u(\hat{u}_2) =s,\quad
\lambda_2^u(\hat{u}_2)=-(s-r-1+\theta),
$$
and
$$
\lambda_1^u(\hat{u}_3)=-(\delta+r)A+s,\quad
\lambda_2^u(\hat{u}_3)=-(s-r-1)A,
$$
where $A=(s-r-1)/\theta+1$.

\begin{figure}
\centering
\includegraphics[width=0.65\textwidth]{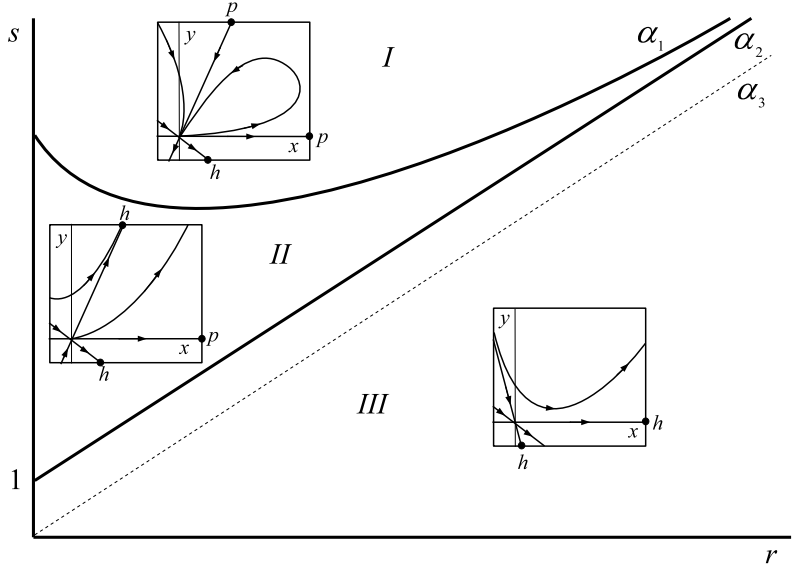}
\caption{Phase-parameter portrait of a neighborhood of the origin
of system \eqref{ex1:1} given as a cut of positive parameter space
for arbitrary fixed values of $0<\theta<1$ and $0<\delta<1$. The
bifurcation boundaries are $\alpha_1\colon
\lambda_1^u(\hat{u}_3)=0$ and $\alpha_2\colon
\lambda_2^u(\hat{u}_1)=0$ ($h$ -- hyperbolic branch, $p$ --
parabolic branch)}
\label{fig:ex1}       
\end{figure}

Using $\lambda_2^u(\hat{u}_1)=0$ and $\lambda_1^u(\hat{u}_3)=0$ as
bifurcation curves we can draw the complete parameter portrait in
$\Omega_+=\Omega\bigcap\mathbb R_+^2$ of the origin (Fig.
\ref{fig:ex1}).

In general, we obtain

\begin{theorem}\label{th_ex1}
Let $0<\delta<1$. For different positive values of the parameters
there exist three types of topologically different structures of
the neighborhood $\Omega_+$ of point $O(0,0)$ of system
\eqref{ex1:1}:

\emph{(i)} An elliptic sector and a part of an attracting
parabolic sector for the parameter values
$\lambda_1^u(\hat{u}_3)>0$ \emph{(}domain $I$ in Figure
\ref{fig:ex1}\emph{)}. The phase curves of the system that tend to
$O(0,0)$ when $t\to-\infty$ are of the form
\begin{equation}\label{ex1:3}
    y=Cx^{(s-r-\delta)/(1-\delta)}(1+o(1));
\end{equation}

\emph{(ii)} A part of a hyperbolic sector and a repelling
parabolic sector for the parameter values
$\lambda_1^u(\hat{u}_3)<0$ and $\lambda_2^u(\hat{u}_1)>0$
\emph{(}domain $II$ in Figure \ref{fig:ex1}\emph{)}. The phase
curves of the system that tend to $O(0,0)$ when $t\to-\infty$ are
of the form \eqref{ex1:3};

\emph{(iii)} A part of a hyperbolic sector for the parameter
values $\lambda_2^u(\hat{u}_1)<0$ \emph{(}domain $III$ in Figure
\ref{fig:ex1}\emph{)}.
\end{theorem}

\textbf{Remark to Theorem \ref{th_ex1}.} In Fig. \ref{fig:ex1} the
line $\lambda_2^u(\hat{u}_2)=0$ is also shown ($\alpha_3$).
Analysis of mutual placing of the branches of the Brouwer sectors
shows that the topological structure of $\Omega_+$ does not change
when we cross $\alpha_3$.
\smallskip

Thus we have proved the existence of an elliptic sector in model
\eqref{ex1:1}, which was overlooked in the original analysis of
Hwang and Kuang \cite{Hwang2}. This dynamical regime corresponds
to the initial growth of the total population size, and the
eventual population extinction can be preceded by a relatively
normal population evolution.

\paragraph{Example 3.}(\textit{Model of Chagas' disease
\emph{\cite{Bussenberg1,{Bussenberg2}}}}) The model of Chagas'
disease from \cite{Bussenberg1,{Bussenberg2}} has the form
\begin{equation}\label{SIex1:1}
\begin{split}
   \frac{dx}{dt} &=(b-r-v)x+(b_1(1-q)+c)y-\beta\frac{xy}{x+y}\,,\\
   \frac{dy}{dt} &=(b_1q-r_1-c)y+vx+\beta\frac{xy}{x+y}\,,
\end{split}
\end{equation}
where all the parameters are nonnegative and $0\leqslant
q\leqslant1$ (see also Section \ref{sec:2}, system \eqref{SIex1}).
After the time change $dt\to(x+y)dt$ we obtain a model in the form
\eqref{mains}, where
\begin{equation*}
    \begin{split}
    P_2(x,y) & =(b-r-v)x^2+(b+b_1-r-b_1q-v-k)xy+(b_1-b_1q)y^2,\\
     Q_2(x,y)   & =vx^2+(b_1q-r_1+v+k)xy+(b_1q-r_1)y^2.
\end{split}
\end{equation*}

In the following we assume that $b-r>0$ and
$\alpha=b-b_1+r_1-r>0$.

The polynomial
$$
F_1(u)=(u+1)(-b_1(1-q)u^2+(k+v-\alpha-b_1(1-q))u+v)
$$
has three roots
$$
\hat{u}_1=-1,\quad
\hat{u}_{\pm}=\frac{A\pm\sqrt{A^2+4vb_1(1-q)}}{2b_1(1-q)},
$$
where $A=k+v-\alpha-b_1(1-q)$. We have $\hat{u}_{-}<0$ and
$\hat{u}_+>0$ if $v>0$. As in the previous example, $\mathbb
R_+^2$ may be contained in two sectors, so we need the branch
characteristics of all the branches:
$$
\lambda_1^u(\hat{u}_1)=k,\quad \lambda_2^u(\hat{u}_1)=\alpha-k,
$$
and
$$
\lambda_1^u(\hat{u}_\pm)=(b_1-r_1)(\hat{u}_{\pm}+1)+\alpha.
$$
We do not explicitly present $\lambda_2^u(\hat{u}_\pm)$, but it is
easy to see the signs of these expressions. We know that $F_1(u)$
is a polynomial of the third order, $F_1(\infty)=-\infty$, and
$F_1(u)$ has one positive and two negative roots. This implies
that $\lambda_2^u(\hat{u}_+)=F'_1(\hat{u}_+)<0$ for any parameter
values. The sign of $\lambda_2^u(\hat{u}_-)$ can be either
negative or positive depending on the sign of
$\lambda_2^u(\hat{u}_1)$ (they have the opposite signs).

Analyzing the branch characteristics we obtain the following
parameter dependent structure of $\Omega_+=\Omega\bigcap\mathbb
R_+^2$ of the origin.

\begin{figure}
\centering
\includegraphics[width=\textwidth]{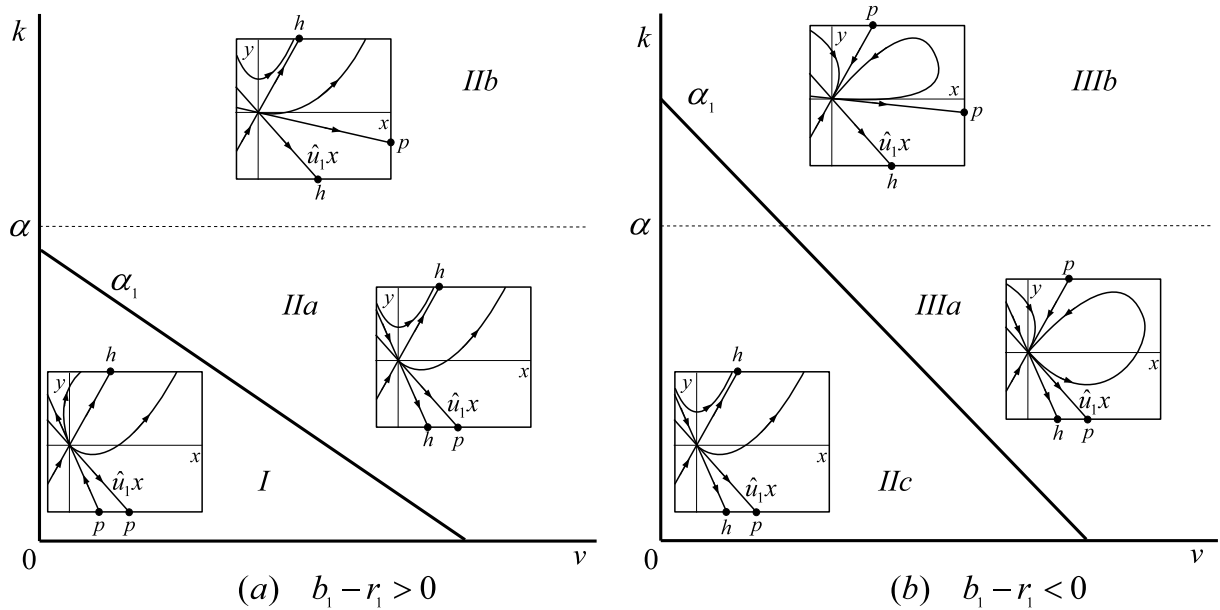}
\caption{Phase-parameter portrait of a neighborhood of the origin
of system \eqref{SIex1:1} given as a cut of positive parameter
space for $(a)$ $b_1-r_1>0$ and $(b)$ $b_1-r_1<0$. The bifurcation
boundary is $\alpha_1=\{(b,\,b_1,\,r,\,r_1,\,k,\,v,\,q)\colon
k=-\frac{\alpha v}{b-r}+\frac{(b_1q-r_1)\alpha}{b_1-r_1}\}$ ($h$
-- hyperbolic branch, $p$ -- parabolic branch)}
\label{fig:ex3}       
\end{figure}

\begin{theorem}\label{th_ex3}
For different positive values of the parameters there exist three
types of topologically different structures of the neighborhood
$\Omega_+$ of the origin of system \eqref{SIex1:1}:

\emph{(i)} A part of a repelling parabolic sector for the
parameter values $b_1-r_1>0$ and $\lambda_1^u(\hat{u}_-)<0$
\emph{(}domain $I$ in Figure \ref{fig:ex3}\emph{)};

\emph{(ii)} A part of a hyperbolic sector and a part of a
repelling parabolic sector for the parameter values $b_1-r_1>0$
and $\lambda_1^u(\hat{u}_-)>0$ or $b_1-r_1<0$ and
$\lambda_1^u(\hat{u}_+)>0$ \emph{(}domain $II$ in Figure
\ref{fig:ex3}\emph{)};

\emph{(iii)} A part of an elliptic sector and a part of an
attracting parabolic sector for the parameter values $b_1-r_1<0$
and $\lambda_1^u(\hat{u}_+)<0$ \emph{(}domain $III$ in Figure
\ref{fig:ex3}\emph{)}.
\end{theorem}

Theorem \ref{th_ex3} gives some additional information about the
behavior of the state variables in comparison with the original
analysis \cite{Bussenberg1}. Most importantly we have found that
population extinction occurs when neighborhood $\Omega$ of
$O(0,0)$ contains an elliptic sector, only part of it belonging to
$\mathbb R_+^2$. Depending on the parameter values, though, this
elliptic sector can be almost entirely in $\mathbb R_+^2$ (domain
$IIIb$ in Fig. \ref{fig:ex3}). In other words, the population
extinction can be essentially non-monotonous, and the total
population size can reach relatively high numbers before
vanishing.

The bifurcation curve $\alpha_1$ corresponds to the presence of a
line of non-isolated equilibrium points of the form
$(x^*,y^*=-(b-r)/(b_1-r_1)x^*)$. In the case $b_1-r_1<0$ this line
belongs to $\mathbb R_2^+$, and the system behavior significantly
changes when $\alpha_1$ is crossed. If parameter values are in
Domain $IIc$ in Fig. \ref{fig:ex3} the population size goes to
infinity; if parameter values are in Domain $III$ the population
becomes extinct with time; while if we are on $\alpha_1$, the line
of non-isolated equilibrium points is an attractor, the asymptotic
size of the population is finite and non-zero, and this asymptotic
size depends on the initial conditions.

It is worth mentioning that our local analysis gives the full
description of the global behavior of system \eqref{SIex1:1} in a
finite part of the phase plane.

Due to Theorem \ref{th3:2} we can add any terms of order 2 or
higher to the right side of \eqref{SIex1:1}, and the parameter
portrait of $\Omega_+$ will not change. For example, it is
reasonable to assume a logistic regulation of the population
\cite{Bussenberg1}.

\section{Conclusions}
In this paper we presented qualitative methods to analyze a wide
class of models of biological populations and communities. These
models are characterized by the presence of complicated singular
equilibria. A significant number of predator-prey, host-parasite,
epidemiological, etc., models, which were recently suggested in
the literature, belong to this class. The main mathematical
peculiarity of these models is their non-analytic structure of the
origin, which is a result of modelling attempts to better reflect
characteristics of the simulated processes at small population
sizes.

Removable singularity peculiar to these models turns into an
analytic equilibrium by a change of independent variable, the
models become well-defined at this equilibrium. However, the
linearization approach fails to infer topological structure of a
small neighborhood of this point, the origin becomes a
non-hyperbolic point, i.e., it has zero eigenvalues for any
parameter values.

Methods of analysis of non-hyperbolic points for two-dimensional
ODE systems with fixed values of the system coefficients were
developed earlier \cite{Berez1,{Bruno},{Dumortier}}. A
neighborhood $\Omega$ of the origin of a smooth second order
system is divided into a finite number of sectors with different
phase behaviors: parabolic, hyperbolic, and elliptic ones. The
parabolic and hyperbolic sectors can be realized in a neighborhood
of a hyperbolic equilibrium, while the elliptic sector is
intrinsic only to a non-hyperbolic equilibrium point.

We presented a complete description of the structure of $\Omega$
and gave asymptotics of $O$-orbits for generic systems whose
Taylor series at the origin start with second order terms. An
exact algorithm is suggested to analyze the structure of the
origin when the system parameters vary. The theory and methods are
illustrated by applying the developed algorithm to the
phase-parameter analysis of some existing models that fall into
the class of ODEs \eqref{mains}. In particular, a model of
anticancer treatment with oncolytic viruses, suggested in
\cite{Novozhilov}, is analyzed in a neighborhood of the origin,
the parameter domain that corresponds to the tumor elimination is
identified; a mathematical model of parasite-host interaction from
\cite{Hwang2} shows that the parasite population can drive the
host to extinction with a preceding population outbreak; a model
of Chagas' disease \cite{Bussenberg1} also demonstrates the effect
of essentially non-monotonous population extinction.

The main attention is paid to the problem of existence and
practical finding of elliptic sectors in the phase plane. An
elliptic sector represents a family of homoclinics to equilibrium,
which means that orbits, starting in this sector, tend to the
origin both for $t\to\infty$ and $t\to-\infty$. Although such type
of dynamic behavior was known in the theory of dynamical systems
for a long time, only relatively recently it was discovered in
biological models. Interpretations of this behavior may be very
fruitful. It can be considered as a specific form of community
extinction accompanied by preceding population outbreaks (see
\cite{Berez5,{Berez6}}).

From the perspective of biological conservation, when population
coexistence is desirable, the parameter domains that are
characterized by the presence of an elliptic sector should be
avoided and their boundaries have to be considered `dangerous',
since both populations go to a rout to extinction after crossing
them. From the perspective of biological control, when population
extinction of one or more populations is desirable, this parameter
area is the most interesting. In this case both populations are
driven to extinction deterministically, an outcome that
prey-dependent models are unable to produce. Let us emphasize that
such peculiarity of the system behavior may be of vital
importance. For example, it is the case in modeling anticancer
therapy with oncolytic viruses. This regime demonstrates a
possibility of complete eradication of the tumor cells
\cite{Novozhilov}.

Similar regimes may exist in more complex and realistic models
with dimensions exceeding $2$. Examples of such systems include
epidemiological models with the number of subpopulations more then
two (e.g., \cite{Bussenberg2,{Hethcote}}), and population models
with three or more trophic levels (e.g., \cite{Hsu1,Hsu3}).

It is our hope that the methods and algorithm presented here could
help not to overlook, as it sometimes happens, the important
regime of deterministic extinction given by the presence of an
elliptic sector.

\section{Appendix}
We consider the vector field
\begin{equation}\label{vecf}
J(x,y)=P(x,y)\frac{\partial}{\partial
x}+Q(x,y)\frac{\partial}{\partial y}
\end{equation}
and the corresponding system of differential equations
\begin{equation}\label{in_sys}
    \frac{dx}{dt} =P(x,y),\quad   \frac{dy}{dt}    =Q(x,y),
\end{equation}
where $$P(x,y)=P_n(x,y)+P^*(x,y),\quad Q(x,y)=Q_n(x,y)+Q^*(x,y).$$
Here $P_n(x,y),\,Q_n(x,y)$ are homogeneous polynomials of the
$n$-th order, and
$$P^*(x,y)=O(|(x,y)|^{n+1}),\quad Q^*(x,y)=O(|(x,y)|^{n+1}).$$

We assume that the origin is an isolated singular point of
\eqref{vecf}: $P(0,0)=Q(0,0)=0$, and let
$F(x,y)=xQ_n(x,y)-yP_n(x,y)$. We also assume that \eqref{vecf} is
non-degenerate, i.e, satisfies
\begin{equation*}
\begin{split}
   (C1)\qquad &\mbox{polynomials $P_n(x,y),\,Q_n(x,y)$ have no common factors of the form}\\
&\mbox{$ax+by$, where at least one the constants $a,\,b$ is
non-zero;}\\
    (C2)\qquad &\mbox{polynomial $F(x,y)$ has no factors of the form $(ax+by)^k$, where $k>1$.}
\end{split}
\end{equation*}

To analyze vector field \eqref{vecf} in a small neighborhood
$\Omega$ of the origin we apply the blowing-up transformations
$(x,y)\rightarrow (x,u)$
\begin{equation}\label{bup1}
x=x,\quad u=y/x,\quad x\neq 0,
\end{equation}
and $(x,y)\rightarrow (v,y)$
\begin{equation}\label{bup2}
y=y,\quad v=x/y,\quad y\neq 0.
\end{equation}

These transformations have the following properties (Ch. 9, Sec.
21 of \cite{Andronov}).
\begin{proposition}\label{pr1}$ $

\emph{(i)} Transformation \eqref{bup1} defines a topological
mapping of the $x$-slit $(x,y)$-plane onto $(x,u)$-plane. Points
of the first (second, third, fourth) quadrant in the $(x,y)$-plane
are mapped respectively onto points of the first (third, second,
fourth) quadrant in the $(x,u)$-plane. The inverse transformation
is defined on the axis $x=0$ of the $(x,u)$-plane, but maps it
onto a single point $(0,0)$ of the $(x,y)$-plane.

\emph{(ii)} Transformation \eqref{bup2} defines a topological
mapping of the $y$-slit $(x,y)$-plane onto $(v,y)$-plane. Points
of the first (second, third, fourth) quadrant in the $(x,y)$-plane
are mapped respectively onto points of the first (second, fourth,
third) quadrant in the $(v,y)$-plane. The inverse transformation
is defined on the axis $y=0$ of the $(v,y)$-plane, but maps it
onto a single point $(0,0)$ of the $(x,y)$-plane.
\end{proposition}

First we apply transformation \eqref{bup1}. After this
transformation and the time change
\begin{equation}\label{ct1}
dt\rightarrow x^{n-1}dt
\end{equation}
we obtain the system
\begin{equation}\label{sys_1}
\begin{split}
\frac{dx}{dt}&=xP_n(1,u)+G_1(x,u),\\
\frac{du}{dt}&=F_1(u)+G_2(x,u),
\end{split}
\end{equation}
where $$F_1(u)=F(1,u),\, G_1(x,u)=P^*(x,ux)/x^{n-1},\,
G_2(x,u)=(Q^*(x,ux)-P^*(x,ux)u)/x^n.$$

Let $\hat{u}$ be a root of $F_1(u)$. Then $(0,\,\hat{u})$ is an
equilibrium point of \eqref{sys_1} with eigenvalues
$P_n(1,\hat{u})$ and $F_1'(\hat{u})$. If vector field \eqref{vecf}
is non-degenerate this equilibrium point is hyperbolic, i.e.,
$P_n(1,\hat{u})F_1'(\hat{u})\neq 0$ (hereafter the prime denotes
the derivative). Summarizing, we obtain

\begin{proposition}\label{pr2}
Equilibrium point $(0,\hat{u})$ of \eqref{sys_1} is a saddle if
$P_n(1,\hat{u})F_1'(\hat{u})< 0$ and a node if
$P_n(1,\hat{u})F_1'(\hat{u})> 0$; this node is a sink if
$P_n(1,\hat{u})<0,\,F_1'(\hat{u})< 0$ and a source if
$P_n(1,\hat{u})>0,\,F_1'(\hat{u})> 0$.
\end{proposition}

According to Proposition \ref{pr1} the point $(x=0,\,y=0)$ is
transformed to $u$-axis in $(x,u)$-plane; $u$-axis consists of the
orbits $\{x=0,\,\hat{u}_i<u<\hat{u}_{i+1}\}$ and of equilibria
$(0,\hat{u}_i)$, where $\hat{u}_i$ are the roots of $F_1(u)$.

Knowledge of the order of the equilibrium points of \eqref{sys_1}
on $u$-axis allows us to infer the structure of the Brouwer
sectors in a small neighborhood $\Omega$ of $O(0,0)$ in
$(x,y)$-plane (except, perhaps, close to the line $x=0$). To
specify possible characteristic directions of $O$-orbits we need
the following theorem (compare to Th. 64 from \cite{Andronov}).

\begin{theorem} \label{th1}$ $

\emph{(i)} Any $O$-orbit of system \eqref{in_sys} is either a
spiral or tends to $O(0,0)$ with a definite tangent, i.e., has a
characteristic direction;

\emph{(ii)} If at least one $O$-orbit is a spiral then all orbits
in some neighborhood of $O(0,0)$ are spirals;

\emph{(iii)} If polynomial $F(x,y)=xQ_n(x,y)-yP_n(x,y)\neq 0$
identically, then coefficients $k$ of all tangent lines $y=kx$ to
$O$-orbits are the roots of the polynomial $F_1(u)=F(1,u)$
\emph{(}except, perhaps, the tangent line $x=0$ corresponding to
the root $v=0$ of polynomial $F_2(v)=-F(v,1)$\emph{)}.
\end{theorem}

The structure of non-degenerate vector field \eqref{vecf} between
the lines, corresponding to neighboring roots of $F_1(u)$, is
described by the following proposition.

\begin{proposition}\label{pr3}
Let $\hat{u}_i<\hat{u}_{i+1}$ be neighboring roots of $F_1(u)$,
i.e., there are no other roots of $F_1(u)$ between them.

\emph{(i)} If points $(0,\hat{u}_i)$, $(0,\hat{u}_{i+1})$ are
saddles, then neighborhood $\Omega$ of the origin of
\eqref{in_sys} has two hyperbolic sectors, one for $x>0$ and
another for $x<0$, whose separatrixes are the curves
$$\alpha_i\colon y=\hat{u}_ix(1+o(x)),\quad \alpha_{i+1}\colon y=\hat{u}_{i+1}x(1+o(1));$$

\emph{(ii)} If points $(0,\hat{u}_i)$, $(0,\hat{u}_{i+1})$ are
nodes, then neighborhood $\Omega$ of the origin of \eqref{in_sys}
has two elliptic sectors, one for $x>0$ and another for $x<0$.
These sectors contain $O$-orbits whose asymptotics are $\alpha_i$
and $\alpha_{i+1}$;

\emph{(iii)} If point $(0,\hat{u}_i)$ is a saddle and point
$(0,\hat{u}_{i+1})$ is a node, then neighborhood $\Omega$ of the
origin of \eqref{in_sys} has two parabolic sectors, one for $x>0$
and another for $x<0$. These sectors contain $O$-orbits whose
asymptotic is $\alpha_{i+1}$.
\end{proposition}

\begin{proof}
The points $(0,\hat{u}_i)$ and $(0,\hat{u}_{i+1})$ are equilibria
of \eqref{sys_1} with eigenvalues
$$\lambda_1^i(\hat{u}_i)=P_n(1,\hat{u}_i),\quad\lambda_2^i=F'_1(\hat{u}_i),$$
and
$$\lambda_1^{i+1}(\hat{u}_{i+1})=P_n(1,\hat{u}_{i+1}),\quad\lambda_2^{i+1}=F'_1(\hat{u}_{i+1}).$$
From the form of the Jacobian matrix it follows that $u=\hat{u}_i$
and $u=\hat{u}_{i+1}$ are the characteristic directions of orbits
tending to $(0,\hat{u}_i)$ and $(0,\hat{u}_{i+1})$ respectively,
i.e., these orbits have asymptotics
$$\gamma_i\colon u=\hat{u}_i(1+o(1)),\quad\gamma_{i+1}\colon
u=\hat{u}_{i+1}(1+o(1)).$$

(i) Let both equilibria be saddles (see, e.g., Fig. \ref{fig:c1},
$i=1$). We have $\lambda_1^i\lambda_2^i<0$,
$\lambda_1^{i+1}\lambda_2^{i+1}<0$, and, due to $(C2)$,
$\lambda_2^i\lambda_2^{i+1}<0$. It implies that
$\lambda_1^i\lambda_1^{i+1}<0$, i.e., the flow along curves
$\gamma_i$ and $\gamma_{i+1}$ has opposite directions (at least
for small $x$). Let point $(x_0,u_0)$, where $u_i<u_0<u_{i+1}$ and
$x_0$ is small enough, belong to orbit $L_0$ of \eqref{sys_1}.
Then $L_0$ approaches $\gamma_{i+1}$ when $t$ increases and
$\gamma_{i}$ when $t$ decreases (Fig. \ref{fig:c1}, $i=1$); $L_0$
forms a hyperbolic-like curve with asymptotics $\gamma_i$ and
$\gamma_{i+1}$. Due to continuity any orbit passing through a
point $(x,u)$ close to $(x_0,y_0)$ is a similar curve. Applying
Proposition \ref{pr1} we obtain a hyperbolic sector in coordinates
$x,y$.

(ii) Let both equilibria be nodes (see Fig. \ref{fig:c2}, $i=2$).
Similarly to the above the flow along curves $\gamma_i$ and
$\gamma_{i+1}$ has opposite directions, one node is a sink and
another one is a source. According to Andronov et al. Ch. 4, Th.
20 \cite{Andronov} infinite number of orbits have asymptotics
$\gamma_i$ and $\gamma_{i+1}$. Taking orbit $L_0$ as above we
obtain that $L_0$ tends to the source for $t\to\infty$ and to the
sink for $t\to-\infty$ forming a parabolic-like curve. The curves
containing points $(x,y)$ close to $(x_0,y_0)$ have the same form.
Applying Proposition \ref{pr1} we obtain an elliptic sector in
coordinates $x,y$.

Case (iii) is considered similarly to the previous ones.
\end{proof}

\textbf{Remark to Proposition \ref{pr3}.} Here it is worth noting
that we also need to specify the direction of the vector field
\eqref{vecf} with respect to the origin. When we speak of the
direction we mean that the flow either goes \textit{towards} the
origin or \textit{from} the origin. For small positive $x$ the
direction of vector field \eqref{vecf} on the line with asymptotic
$y=\hat{u}x(1+o(1))$ coincides with the direction of the vector
field corresponding to \eqref{sys_1} on $u=\hat{u}(1+o(1))$ (i.e.,
either both flows go towards corresponding equilibria or from
them); for small negative $x$ these directions are the same if $n$
is odd and opposite if $n$ is even. This simple fact follows from
the time change \eqref{ct1} and should not be forgotten when
analyzing the direction of the vector field.
\smallskip

The number of equilibria of \eqref{sys_1} of the form
$(0,\hat{u})$ is equal or less than $n+1$. Denote $n_s$ and $n_n$
the numbers of saddles and nodes respectively.

\begin{proposition}\label{pr5}
$$n_s\leqslant n+1,\qquad n_n<n+1.$$
\end{proposition}
\begin{proof}
Let $\hat{u}_i,\,\hat{u}_{i+1}$ be neighboring roots of $F_1(u)$
and $\lambda^i_1=P_n(1,\hat{u}_i),\,\lambda_2^i=F_1'(\hat{u}_i)$
are the eigenvalues of equilibrium $(0,\hat{u}_i)$. Due to $(C2)$
we have that $\lambda_2^i\lambda_2^{i+1}<0$. The polynomial
$P_n(1,u)$ can change sign at most $n$ times. Polynomials $F_1(u)$
and $P_n(1,u)$ have opposite signs for large $u$. We have to show
that the sequence
$\{\lambda_1^1\lambda_2^{1},\ldots,\lambda_1^{n+1}\lambda_2^{n+1}\}$
cannot have all the elements positive. If, e.g.,
$F_1(\infty)=+\infty$ then $F_1'(\hat{u}_{n+1})>0$, where
$\hat{u}_{n+1}$ is the largest root of $F_1(u)$. If we suppose
that $P_n(1,\hat{u}_{n+1})>0$, it means that $P_n(1,u)$ has
already changed its sign at least once and we have $n-2$ possible
sign changes for the rest $n$ roots. Here is a contradiction, and
we cannot have $n_n=n+1$. Similar arguing shows that the sequence
$\{\lambda_1^1\lambda_2^{1},\ldots,\lambda_1^{n+1}\lambda_2^{n+1}\}$
can have all the elements negative.
\end{proof}

Blowing-up transformation \eqref{bup1} allows us to explore the
behavior of vector field \eqref{vecf} anywhere except for the axis
$x=0$. The behavior of \eqref{vecf} close to $y$-axis can be
investigated with the help of the second blowing up transformation
\eqref{bup2}.

After the time change
\begin{equation}\label{ct2}
dt\rightarrow y^{n-1}dt
\end{equation}
we obtain the system
\begin{equation}\label{sys_2}
\begin{split}
\frac{dv}{dt}&=F_2(v)+H_1(v,y),\\
\frac{dy}{dt}&=yQ_n(v,1)+H_2(v,y),
\end{split}
\end{equation}
where
$$F_2(v)=-F(v,1),\,H_1(v,y)=(P^*(vy,y)-Q^*(vy,y)v)/y^n,\,H_2(v,y)=Q^*(vy,y)/y^{n-1}.$$

According to Proposition \ref{pr1} systems \eqref{in_sys} and
\eqref{sys_2} are equivalent in all the points $y\neq 0$; the
point $(x=0,y=0)$ in transformed into axis $v$ in $(v,y)$-plane;
the axis $v$ consists of the orbits
$\{\hat{v}_i<v<\hat{v}_{i+1},\,y=0\}$ and of equilibria
$(\hat{v}_i,\,0)$, where $\hat{v}_i$ are the roots of $F_2(v)$.

There is a strict correspondence between the number and types of
equilibrium points of \eqref{sys_1} on $u$-axis, and the number
and types of equilibrium points of \eqref{sys_2} on $v$-axis.

\begin{proposition}\label{pr6}
Let vector field \eqref{vecf} be non-degenerate, polynomial
$F_1(u)$ have a root $\hat{u}\neq 0$, and
$\lambda_1^u,\,\lambda_2^u$ be the eigenvalues of the singular
point $(0,\hat{u})$ of system \eqref{sys_1}. Then

\emph{(i)} polynomial $F_2(v)$ has a root $\hat{v}=1/\hat{u}$;

\emph{(ii)} the eigenvalues $\lambda_1^v,\,\lambda_2^v$ of the
singular point $(\hat{v},0)$ of system \eqref{sys_2} satisfy the
relations
$\lambda_1^v=\lambda_1^u/\hat{u}^{n-1},\,\lambda_2^v=\lambda_2^u/\hat{u}^{n-1}$.
\end{proposition}
\begin{proof}

(i) We have $F_1(u)=F(1,u),\,F_2(v)=-F(v,1)$ and the equalities
$$F(x,ux)=x^{n+1}F_1(u),\,F(vy,y)=-y^{n+1}F_2(v) .$$
$F_1(\hat{u})=0$ implies $F(x,x\hat{u})=0$ for any $x$. Consider
$$F(\hat{v}y,\hat{u}\hat{v}y)=\hat{u}^{n+1}\hat{v}^{n+1}y^{n+1}F(1/\hat{u},1)=0,$$
which means that $1/\hat{u}$ is a root of $F_2(v)$.

(ii) We use the following notations:
$$
\lambda_1^u=P_n(1,u),\quad\lambda_2^u=F'_1(u),\quad\lambda_1^v=Q_n(v,1),\quad\lambda_2^v=F'_2(v).
$$

We have $F(1,\hat{u})=-\hat{u}P_n(1,\hat{u})+Q_n(1,\hat{u})$.
Thus,
$\lambda_1^u=P_n(1,\hat{u})=Q_n(1,\hat{u})/\hat{u}=\hat{u}^nQ_n(1/\hat{u},1)/\hat{u}=\hat{u}^{n-1}Q_n(1/\hat{u},1)=\hat{u}^{n-1}\lambda_1^v$.

Next, we can write $F(1,u)=u^{n+1}F(1/u,1)$ for any $u\neq 0$.
Taking the derivatives of this relation and putting $u=\hat{u}$ we
obtain
$$
\lambda_2^u=F_u'(1,\hat{u})=(n+1)\hat{u}^nF(1/\hat{u},1)-\left.\hat{u}^{n-1}F_v'(v,1)\right|_{v=1/\hat{u}}=\hat{u}^{n-1}\lambda_2^v.
$$
\end{proof}

\begin{corollary}\label{cor1}
The topological type of equilibrium point $(0,\hat{u})$,
$\hat{u}\neq 0$, of system \eqref{sys_1} coincides with the
topological type of equilibrium point $(1/\hat{u},0)$ of system
\eqref{sys_2}, that is, both points are saddles or nodes
simultaneously.
\end{corollary}

The only root of $F_2(v)$ that cannot be described with the
knowledge of types of the roots of $F_1(u)$ is, due to Proposition
\ref{pr6}, $v=0$.

\begin{proposition}\label{pr7}
Let system \eqref{in_sys} satisfy $(C1)$ and $(C2)$ and have an
isolated singular point at the origin. Then the point $(v=0,y=0)$
is an equilibrium of \eqref{sys_2} if and only if $F_2(v)$ has a
root $v=0$. It is a saddle if $F_2'(0)Q_n(0,1)<0$ and a node if
$F_2'(0)Q_n(0,1)>0$; this node is a sink if
$F_2'(0)<0,\,Q_n(0,1)<0$ and a source if $F_2'(0)>0,\,Q_n(0,1)>0$.
\end{proposition}

Obviously, an analogue of Proposition \ref{pr3} holds for
blowing-up transformation \eqref{bup2}. The only thing that should
be emphasized here is the simultaneous directions of vector field
\eqref{vecf} and the vector field corresponding to \eqref{sys_2}
(see the remark to Proposition \ref{pr3}).

The lines $y=\hat{u}_ix$ and $x=\hat{v}_iy$, where $\hat{u}_i$ and
$\hat{v}_i$ are the roots of polynomials $F_1(u)$ and $F_2(v)$
respectively, divide a neighborhood of the origin of
\eqref{in_sys} into sectors, whose boundaries are the branches of
these lines (each line consists of two branches).

Below in the proof of Proposition \ref{pr} we use the notations
introduced in Proposition \ref{pr6}.

We shall call the branches, corresponding to the line $y={u}_ix$,
of a sector in phase space $(x,y)$ \textit{hyperbolic} if for the
eigenvalues of the equilibrium $(0,\hat{u}_i)$  the inequality
$\lambda^u_{1}(\hat{u}_i)\lambda^u_2(\hat{u}_i)<0$ holds, and
\textit{parabolic} if
$\lambda^u_{1}(\hat{u}_i)\lambda^u_2(\hat{u}_i)>0$ holds. An
analogous definition applies to the branches corresponding to the
line $x=\hat{v}_iy$.

\begin{proof}[Proof of Proposition \ref{pr}]
If a sector belongs to the half plane $x>0$ (this also implies
that there is a complementary sector for $x<0$) or $y>0$, the
assertion follows from Proposition \ref{pr3} or from its analogue
for the blowing-up transformation \eqref{bup2}. There are two
cases which are not covered by Proposition \ref{pr3}:

(i) A sector is not contained in half plane $x>0$, as well as in
$y>0$ (e.g., all the roots of $F_1(u)$ satisfy the condition
$\hat{u}_i>0$);

(ii) A sector coincides with the first (or second) quadrant of the
phase plane, i.e., $\hat{u}=0$ and $\hat{v}=0$ are the roots of
polynomials $F_1(u)$ and $F_2(v)$.

We start with the case (i).

Due to the assumptions we have that $x=0$ is not a characteristic
direction and the sector under consideration is confined by the
branches of lines corresponding to the least and the largest roots
of $F_1(u)$, which we denote $\hat{u}$ and $\hat{U}$.

The main idea of the proof is exactly the same as the one used in
Proposition \ref{pr3}. We need to show that the vector field is
coordinated. The required statement then follows from the
continuity of the vector field.

Let $n$ be even. The polynomial $F_1(u)$ can have $2k+1$ roots,
which implies that $\lambda_2^u(\hat{u})\lambda_2^u(\hat{U})>0$.
In the proof of Proposition \ref{pr3} we had the opposite
inequality for two neighboring roots, but considered the sectors,
both branches of which lay in the half plane $x>0$. Due to the
time change \eqref{ct1} for even $n$ the direction of the vector
field of system \eqref{in_sys} on the asymptotic $y=\hat{u}x$ is
opposite to the direction of the vector field of system
\eqref{sys_1} on the asymptotic $u=\hat{u}$ for $x<0$. Repeating
the proof of Proposition \ref{pr3} and taking into account the
direction of the vector fields (e.g., if both roots correspond to
saddles of system \eqref{sys_1} this yields that
$\lambda_1^u(\hat{u})\lambda_1^u(\hat{U})>0$ and the vector field
on the branches of the sector has opposite directions in respect
to the origin) we prove the claim.

The case of only one root of $F_1(u)$ is easily included in the
above reasoning.

If $n$ is odd, polynomial $F_1(u)$ has $2k$ roots. We assume here
that $k>0$ (see also general remarks at the end of Appendix). It
implies that $\lambda_2^u(\hat{u})\lambda_2^u(\hat{U})<0$. Now the
direction of the vector field on the branch $\hat{u}x$ coincides
with the direction of the vector field on the line $u=\hat{u}$ of
system \eqref{sys_1}, and using the reasoning from Proposition
\ref{pr3} we prove the statement.

(ii) We consider the case when a sector coincides with the first
quadrant. The axes are asymptotics to the orbits starting in the
first quadrant because there are no other asymptotics of the form
$y=\hat{u}_ix$, where $\hat{u}_i>0$. From Proposition \ref{pr1} it
follows that systems \eqref{sys_1} and \eqref{sys_2} are
equivalent inside the first quadrant and we can consider them
simultaneously.

We have that $\hat{u}=0$ is a root of $F_1(u)$ and $\hat{v}=0$ is
a root of $F_2(v)$. This and non-degeneracity of the vector field
\eqref{vecf} mean that Taylor series of $F_1(u)$ and $F_2(v)$
start from the terms of order $n$.

If $n$ is even, then $F_1(u)$ can have $2k$ roots, one of them is
zero. If we denote the least root of $F_1(u)$ as $\hat{u}$, we
obtain $\lambda_2^u(0)\lambda_2(\hat{u})<0$. The root
$\hat{v}=1/\hat{u}$ is the largest root of $F_2(v)$ (except for
zero), and, from Proposition \ref{pr6}, we have
$\lambda_2^u(\hat{u})\lambda_2^v(\hat{v})<0$. This yields that
$\lambda_2^u(0)\lambda_2^v(0)<0$.

If $n$ is odd, then again similar arguing shows that
$\lambda_2^u(0)\lambda_2^v(0)<0$. The rest of the proof repeats
the proof of Proposition \ref{pr3}.
\end{proof}

Now we are ready to prove the basic theorem.

\begin{proof}[Proof of Theorem \ref{th3:2}] We have $n=2$ and polynomial
$F_1(u)$ can have three, two, one or no real roots.

First we consider the case of three real roots of $F_1(u)$. Note
that we can include the case when one of the roots is zero (in
this case, due to Proposition \ref{pr6}, $F_2(v)$ has two non-zero
roots). It is possible, generally speaking, seven cases of mutual
order of the equilibrium points of \eqref{sys_1} on $u$-axis: (1)
three saddles; (2) node-node-saddle; (3) node-saddle-node; (4)
saddle-node-node; (5) saddle-saddle-node; (6) saddle-node-saddle;
(7) node-saddle-saddle. The case when there are three nodes cannot
be realized due to Proposition \ref{pr5}.

Placing the Brouwer sectors between the neighboring roots is
accomplished with the help of Proposition \ref{pr}.

\begin{figure}
\centering
\includegraphics{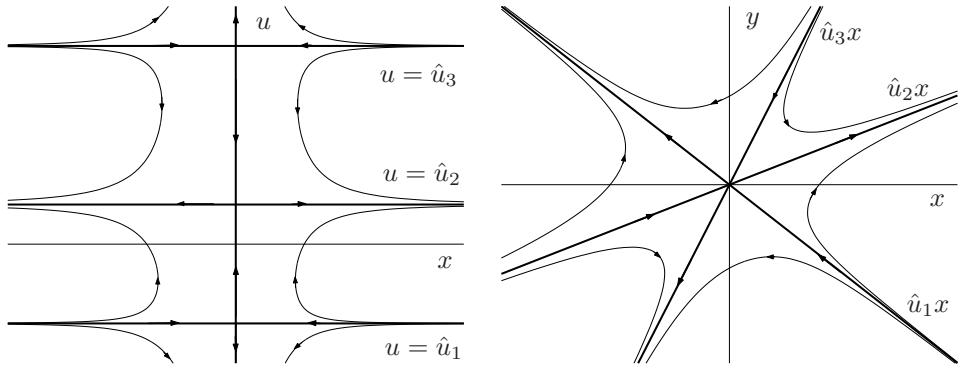}
\caption{Correspondence between the phase plane $(x,u)$ of system
\eqref{sys_1} and neighborhood $\Omega$ of the origin of system
\eqref{in_sys}. The case of six hyperbolic sectors}
\label{fig:c1}       
\end{figure}

\begin{figure}
\centering
\includegraphics{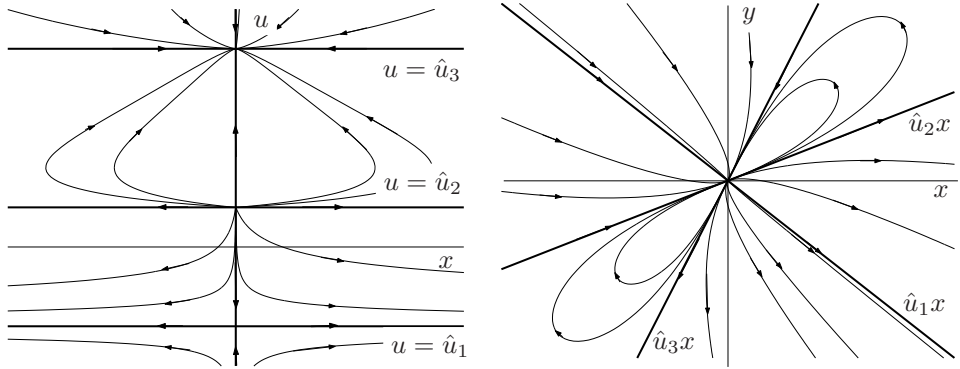}
\caption{Correspondence between the phase plane $(x,u)$ of system
\eqref{sys_1} and neighborhood $\Omega$ of the origin of system
\eqref{in_sys}. The case of two parabolic and two elliptic
sectors}
\label{fig:c2}       
\end{figure}

Thus we have that case (1) corresponds to 6 hyperbolic branches
and 6 hyperbolic sectors (Fig. \ref{fig:c1}); cases (2)-(4)
correspond to two hyperbolic and four parabolic branches or,
respectively, two elliptic sectors and two parabolic sectors (Fig.
\ref{fig:c2}), here four parabolic sectors described by
Proposition \ref{pr} merge into two; cases (5)-(7) correspond to
four hyperbolic and two parabolic branches, or to two hyperbolic
and two parabolic sectors (Fig. \ref{fig:c3}).

\begin{figure}
\centering
\includegraphics{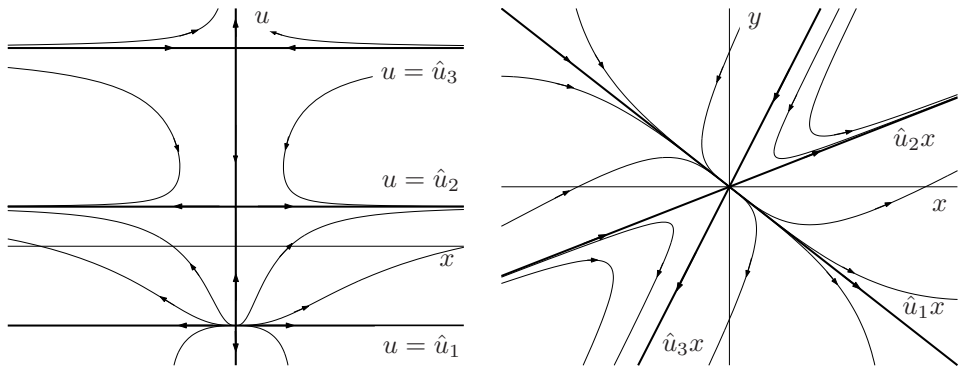}
\caption{Correspondence between the phase plane $(x,u)$ of system
\eqref{sys_1} and neighborhood $\Omega$ of the origin of system
\eqref{in_sys}. The case of two parabolic and two hyperbolic
sectors}
\label{fig:c3}       
\end{figure}

Now let $F_1(u)$ have only one real root. We can include the case
$\hat{u}=0$ due to $(C1)$ and $(C2)$ ($F_2(v)$ has no real roots
in this case). This root can correspond to a saddle or node
equilibrium point $(0,\hat{u})$ of \eqref{sys_1}. The first case
agrees (Proposition \ref{pr}) with the presence of two hyperbolic
sectors (Fig. \ref{fig:c4}); the second one agrees with the
presence of two elliptic sectors (Fig. \ref{fig:c5}).

\begin{figure}
\centering
\includegraphics{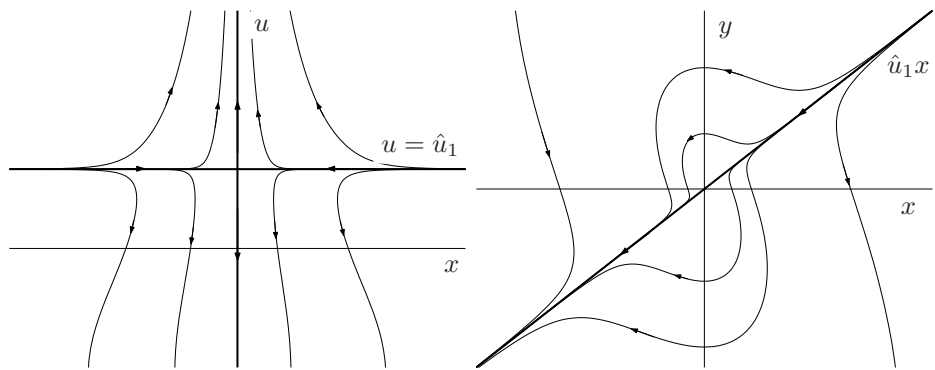}
\caption{Correspondence between the phase plane $(x,u)$ of system
\eqref{sys_1} and neighborhood $\Omega$ of the origin of system
\eqref{in_sys}. The case of two hyperbolic sectors}
\label{fig:c4}       
\end{figure}

\begin{figure}
\centering
\includegraphics{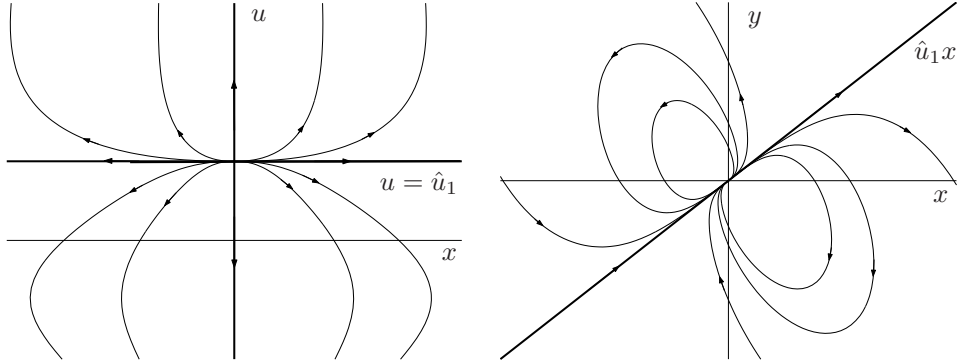}
\caption{Correspondence between the phase plane $(x,u)$ of system
\eqref{sys_1} and neighborhood $\Omega$ of the origin of system
\eqref{in_sys}. The case of two parabolic and two elliptic
sectors}
\label{fig:c5}       
\end{figure}

If polynomial $F_1(u)$ has two real roots which are not equal to
zero, it follows from $(C1)$, $(C2)$ and Proposition \ref{pr6}
that $F_2(v)$ has three real roots and we should consider $F_2(v)$
instead of $F_1(u)$ (analogous to the case of three real roots of
$F_1(u)$).

Polynomial $F_1(u)$ can have two real roots, and one of them
$\hat{u}=0$. This necessarily implies (together with Proposition \eqref{pr6}) that $F_2(v)$ also
has two real roots, one of them $\hat{v}=0$. So we have three
roots ($\hat{u}\neq0$ corresponds to $\hat{v}=1/\hat{u}$ which of
the same type). Simple arguing shows that these three roots can be
(1) three saddles (six hyperbolic sectors); (2) two nodes and a
saddle (two elliptic and two parabolic sectors); (3) a node and
two saddles (two parabolic and two hyperbolic sectors).

If $F_1(u)$ has no real root it implies that $F_2(v)$ has the only
root $\hat{v}=0$ and this case is similar to the case of one real
root of $F_1(u)$: two hyperbolic or two elliptic sectors depending
on the type of $(0,\hat{v}=0)$.

Other cases are not possible in system \eqref{in_sys} with $n=2$
that satisfies $(C1)$ and $(C2)$, which completes the proof.
\end{proof}

If we specify some additional conditions on \eqref{vecf}, we can
say more about asymptotics to the lines $y=0$ and $x=0$ of vector
field \eqref{vecf}. Namely, if
$$
(C3)\qquad P_n(0,y)\equiv0\Rightarrow P^*(0,y)\equiv 0,\quad
Q_n(x,0)\equiv0\Rightarrow Q^*(x,0)\equiv 0
$$
hold, we can prove

\begin{proposition}\label{pr4}
Let $y=0$ consist of orbits of non-degenerate vector field
\eqref{vecf} satisfying $(C3)$, and $\beta=q_{n,1}/{p_{n,1}}>1$.
Then \eqref{vecf} has a family of $O$-orbits
$y=c|x|^{\beta}(1+o(1))$, where $c$ is an arbitrary constant.
\end{proposition}
\begin{proof}
If $y=0$ consists of orbits of vector field \eqref{vecf} then
$\hat{u}=0$ is a root of
$F_1(u)=-p_{0,n+1}u^{n+1}+\ldots+(q_{n,1}-p_{n,1})u+q_{n+1,0}$. It
implies that $q_{n+1,0}=0$, and $y=0$ is a factor of $Q_n(x,y)$,
from which $Q^*(x,0)\equiv 0$ follows (due to $(C3)$). From $(C1)$
it follows that $p_{n,1}\neq 0$. According to Proposition
\ref{pr2} the hyperbolic equilibrium point $(x=0,u=0)$ is a node
because we assumed that $q_{n,1}/{p_{n,1}}>1$ which means
$F_1'(0)P_n(1,0)=p_{n,1}(q_{n,1}-p_{n,1})>0$. There is a family of
characteristic orbits of the form $u=c|x|^{\beta-1}(1+o(1))$ where
$c$ is an arbitrary constant. Returning to variables $x,\,y=ux$
completes the proof.
\end{proof}

Using the second blow-up transformation \eqref{bup2} we obtain

\begin{proposition}\label{pr8}
Let $x=0$ consist of orbits of non-degenerate vector field
\eqref{vecf}, and $\beta=p_{1,n}/{q_{1,n}}>1$. Then vector field
\eqref{vecf} has a family of $O$-orbits $x=c|y|^{\beta}(1+o(1))$,
where $c$ is an arbitrary constant.
\end{proposition}
The proof of Proposition \ref{pr8} is similar to the proof of
Proposition \ref{pr4}.

\begin{proof}[Proof of Theorem \ref{th3:3}] The proof follows from Theorem
\ref{th1} and Propositions \ref{pr4} and \ref{pr8}.
\end{proof}

\paragraph{General remarks.} Vector field \eqref{vecf} is a particular case of so-called
vector field with a fixed Newton diagram (e.g., \cite{Arnold}).
The conditions of existence of orbits with characteristic
directions for vector fields with an arbitrary Newton diagram are
also known \cite{Berez2,Berez3,Berez4}. The following facts are
stated for a homogeneous case but can be carried onto more general
situations almost literally.

Let
$$
J_n=P_n(x,y)\frac{\partial}{\partial
x}+Q_n(x,y)\frac{\partial}{\partial y}
$$
be a homogeneous vector field having characteristic directions.
The following theorem holds.
\begin{theorem}
There exists a neighborhood $\Omega$ of singular point $O(0,0)$,
such that vector field \eqref{vecf} is topologically equivalent to
vector field $J_n$ in $\Omega$.
\end{theorem}

In case when $J_n$ has no orbits with characteristic directions
point $O(0,0)$ is monodromic (this can be the case only if $n$ is
odd). If we assume that
$$
(C4)\quad\mbox{Polynomial $F(x,y)$ has no multiple complex
factors}
$$
holds, then the cases of focus or center can be identified.

Denote
$$
l=\sum_{\textrm{Im}\, z>0}\textrm{res}_z\frac{P_n(1,u)}{F(1,u)}
$$
the generalized first Lyapunov value.
\begin{theorem}
Let $O(0,0)$ be a monodromic singular point of non-degenerate
vector field $J_n$ satisfying $(C4)$. Then $O(0,0)$ is a focus if
$l\neq 0$. This focus is attracting if $l<0$ and repelling if
$l>0$.
\end{theorem}

Thus, in principle, it is possible to analyze non-hyperbolic
equilibrium $O(0,0)$ of system \eqref{in_sys} for an arbitrary
$n$.

\paragraph{Acknowledgements.} The work of FSB has been supported in part by NSF Grant
\#634156 to Howard University.

\end{document}